\documentclass[letterpaper,onecolumn,nofootinbib]{revtex4}

\usepackage{graphicx}
\usepackage{amsfonts}
\usepackage{amsmath}
\usepackage{amssymb}
\usepackage{caption}
\usepackage{subcaption}
\usepackage{footnote}
\usepackage{footmisc}
\usepackage{color}
\usepackage{comment}

\providecommand{\U}[1]{\protect\rule{.1in}{.1in}}
\newtheorem{theorem}{Theorem}

\newtheorem{definition}[theorem]{Definition}

\newenvironment{proof}[1][Proof]{\noindent\textbf{#1.} }{\ \rule{0.5em}{0.5em}}

\definecolor{nblue}{rgb}{0.2,0.2,0.7}
\definecolor{ngreen}{rgb}{0.2,0.6,0.2}
\definecolor{nred}{rgb}{0.7,0.2,0.2}
\definecolor{nblack}{rgb}{0,0,0}

\definecolor{dred}{rgb}{.8,0.2,.2}
\definecolor{ddred}{rgb}{.8,0.5,.5}
\definecolor{dblue}{rgb}{.2,0.2,.8}

\def\indep{\perp\!\!\!\!\perp}

\begin{document}

\title{The lesson of causal discovery algorithms for quantum correlations: \\ Causal explanations of Bell-inequality violations require fine-tuning}
\author{Christopher J. Wood}
\affiliation{Institute for Quantum Computing, Waterloo, Ontario, Canada, N2L 3G1}
\affiliation{Department of Physics and Astronomy, University of Waterloo, Waterloo, Ontario, Canada N2L 3G1}
\author{Robert W. Spekkens}
\affiliation{Perimeter Institute for Theoretical Physics, Waterloo, Ontario, Canada N2L
2Y5}
\date{\today}

\begin{abstract}
An active area of research in the fields of machine learning and statistics is the
development of \emph{causal discovery algorithms}, the purpose of which is
to infer the causal relations that hold among a set of variables from the
correlations that these exhibit. We apply some of these algorithms to the
correlations that arise for entangled quantum systems.  We show that they
cannot distinguish correlations that satisfy Bell inequalities from
correlations that violate Bell inequalities, and consequently that they cannot do
justice to the challenges of explaining certain quantum correlations
causally.  Nonetheless, by adapting the conceptual tools of causal inference, we can show that any attempt to provide a causal explanation of nonsignalling correlations that violate a Bell inequality must contradict a core
principle of these algorithms, namely, that an observed statistical
independence between variables should not be explained by fine-tuning of the
causal parameters.  In particular, we demonstrate the need for such fine-tuning for most of the causal mechanisms that have been proposed to underlie Bell correlations, including superluminal causal influences, superdeterminism (that is, a denial of freedom of choice of settings), and retrocausal influences which do not introduce causal cycles.
\end{abstract}

\maketitle

\section{Introduction}
\label{sec:intro}

A causal relation, unlike a correlation, is an asymmetric relation that can
support inferences about the consequences of interventions and about
counterfactuals. The sun rising and the rooster crowing are strongly
correlated, but to say that the first is the cause of the second is to say
more. In particular, it says that forcing the rooster to crow early will not
precipitate an early dawn, whereas causing the sun to rise early (for
instance, by moving the rooster eastward), can lead to some early crowing.
Nonetheless, causal structure has implications for the observed correlations
and consequently one can make inferences about the causal structure based on
the observed correlations. Indeed, there has been much progress in the
last twenty-five years on how to make such inferences, progress that has
been primarily due to philosophers and researchers in the field of machine
learning and which is well summarized in the books of Pearl~\cite{Pearl2009} and of Spirtes, Glymour
and Scheines (SGS) \cite{Spirtes2001}. Such inference schemes are known
as \emph{causal discovery algorithms}. In this article, we shall consider
the question of what some prominent causal discovery algorithms have to say about
the causal structure that might underlie quantum correlations, in particular
those that violate Bell inequalities.

Suppose that one conducts measurements on a pair of systems that have been
prepared together, and then removed to distant locations such that the
outcome at each wing of the experiment is outside the future light cone of
the measurement choice in the other wing. Suppose further that one finds that the
correlations so obtained violate Bell inequalities.  If one insists on a \emph{causal} explanation of these correlations, then it would seem that one must admit that the causes must propagate faster than the speed of light. But this is in tension with the fact that one cannot send signals faster than the speed of light. We take this tension to be the mystery of Bell's theorem: if there are indeed
superluminal causes, then why can't we use them to send superluminal
signals? In this article, we will show that the principles behind causal
discovery algorithms can clarify the nature of this tension.  We also show that this tension persists in more exotic proposals for achieving a causal explanation of Bell inequality violations such as superdeterminism, which is an assumption that at least one of the measurement settings is influenced by a variable that is a common cause of the outcome on the opposite wing (and hence this setting variable is not freely chosen), and retrocausation, wherein causes propagate counter to the standard direction of time.

We consider the most prominent causal discovery algorithms, which take as their input the set of conditional independence relations that hold among the observed variables.  No other feature of the probability distribution is relevant for them.\footnote{These algorithms also look only at \emph{statistical} independences.  They do not consider \emph{algorithmic} independences, which have recently been shown to encode information about the causal structure.  In particular, Sun \emph{et al.}\cite{Sun} have shown that one can make use of the fact that the conditional distribution of an effect given its causes is typically a smoother distribution than the marginal distribution of the effect, and the marginal distribution of a cause is typically a smoother distribution than the conditional distribution of the cause given some of its effects.}
Our analysis will reveal that such algorithms do not capture the insights of Bell's theorem.  It follows that there is an opportunity for researchers in the field of quantum foundations with expertise on Bell's theorem to improve upon existing causal discovery algorithms.   
Indeed, in the time since a preprint of this article first appeared, the process has already begun. 
Inspired by entropic Bell inequalities and building on the work of Ref.~\cite{FritzChaves}, it has recently been shown in Ref.~\cite{Chavesetal} that the causal structure implies certain entropic inequalities on the joint probability distribution.  We anticipate that there are many more opportunities for improvements to causal inference 
based on ideas from the field of quantum foundations.\footnote{Other work in the field of machine learning has appealed to statistical features besides conditional independence relations, but not the features of correlations that are relevant for Bell's theorem.  Peters \emph{et al.}\cite{Peters} demonstrate that if one is promised an additive noise model, then features of the joint distribution can often distinguish cause from effect in the case of a  distribution on a pair of variables, where there are no conditional independence relations to guide the analysis.}

The distinction between causal and inferential concepts is an instance of
the distinction between ontic concepts (those pertaining to reality) and
epistemic concepts (those pertaining to our knowledge of reality).  Within
the field of statistics, disentangling causal and inferential concepts is
notoriously difficult and controversial, as is the question of when causal
claims are supported by the observed correlations. In the quantum realm,
where there is even less agreement about which parts of the formalism refer
to ontic concepts and which refer to epistemic concepts, the problem is
compounded~\cite{Harrigan2010}.  As such, we shall try to present our analysis in
a manner that does not presume any particular interpretation of quantum
theory.  For instance, given that different interpretations disagree on
whether quantum theory implies an objective indeterminism in nature or not,
we shall not presume any particular answer to this question.  Instead, we
simply focus on the operational predictions of the theory.

Some previous work has already considered Bell's theorem from the perspective of causal discovery algorithms.
In particular, the books by Pearl  \cite{Pearl2009} and by SGS \cite{Spirtes2001} comment briefly on the question. They both assert that Bell's theorem forces a dilemma between (i) abandoning a particular notion of locality, that there are no superluminal causal influences, and  
(ii) abandoning the assumption that if two variables are statistically dependent, then this is explained either by the existence of a cause from one to the other or a common cause acting on both, or a combination of the two mechanisms.  Assumption (ii) underlies what is called the ``causal Markov condition'', but we will refer to it here simply as \emph{Reichenbach's principle}; in a slogan, it asserts that correlations must be explained causally.  \color{black}
One can legitimately quibble with 
the claim that Bell's theorem forces such a dilemma
on the grounds that there are other assumptions that go into the theorem:  the absence of {\em superdeterminism} (an assumption that is often characterized as the existence of freedom in the choice of settings), and the absence of retrocausal influences, for instance.  
Nonetheless, this is an improvement over the standard characterization of Bell's theorem as forcing a dilemma between abandoning locality and abandoning \emph{realism}.  It has always been rather unclear what precisely is meant by ``realism''.  Norsen has considered various philosophical notions of realism and concluded that none seem to have the feature that one could hope to save locality by abandoning them~\cite{norsen2007against}.  For instance, if realism is taken to be a commitment to the existence of an external world, then the notion of locality -- that every causal influence between physical systems propagates subluminally -- already \emph{presupposes} realism.
Furthermore, we will show that the tools of causal inference can also be used to argue for the implausibility of superdeterminism and retrocausal influences.

Our first conclusion is a relatively straightforward one.  We note that in the case of  a Bell scenario, where a pair of systems is prepared together then separated and each subjected to a measurement, \color{black} all correlations exhibit the following conditional independence relations among the observable variables:
\begin{enumerate}
\item Marginal independence of the measurement setting variables,
\item No-signalling, that is, conditional independence of the outcome at one wing of the experiment and the setting at the opposite wing given the setting at the first wing.
\end{enumerate}
Except for independences that are due to special degeneracies in the quantum state,  these are the \emph{only} conditional independences arising in Bell scenarios.
These independences characterize both the correlations that satisfy all the Bell inequalities, and the correlations that violate some Bell inequality. Therefore, if the causal discovery algorithm takes as its input not the full distribution but only the conditional independence relations that hold in the distribution (as is the case with the prominent such algorithms), then this algorithm cannot distinguish
correlations that violate Bell inequalities from correlations that satisfy
Bell inequalities. The input to such algorithms is simply too impoverished
to see the difference. It follows that the causal distinctions that \emph{do}
exist between these correlations --- those that are implied by Bell's theorem --- cannot be
recognized by these algorithms.  They may consequently make incorrect
assessments of what causal structure is implied by a given set of
correlations.

By explicitly applying the standard causal discovery algorithms to the conditional independence relations that characterize a Bell scenario, we draw attention to the fact that the output of such algorithms must be interpreted with great care, lest one be led to an incorrect conclusion about the viability of certain causal explanations.  We look at both the case where one presumes that the settings and outcomes are the only causally relevant variables, i.e., the case of no hidden variables, and the case where one allows hidden variables.
 
Finally, we set aside the details of existing algorithms and consider simply what the core principles underlying these algorithms imply about the possibility of causal explanations of Bell inequality violations. 
We demonstrate that any causal model that can hope to explain Bell-inequality-violating correlations (or even to explain Bell-inequality-satisfying correlations without recourse to hidden variables) has the feature that in order to explain the conditional independencies among the observed variables, in particular the no-signalling constraints, it must involve a \emph{fine-tuning} of the causal parameters.

So, in the end, we obtain a characterization of Bell's theorem that is quite far from its standard characterization as a denial of ``local realism''.  
The assumptions that go into this new characterization are: the framework of causal models, which incorporates Reichenbach's principle that correlations should be explained causally, as well as the principle that conditional independence relations should not be explained by fine-tuning.  As we shall see, the no fine-tuning principle, applied to the observed independences in a Bell scenario (including the lack of superluminal signals), implies the lack of superluminal causal influences, which is Bell's notion of local causality.  So Bell's notion of local causality is {\em derived} as one particular consequence of no fine-tuning in this approach.  The real innovation of this approach, however, is that the no fine-tuning principle together with the observed indepedences also rule out superdeterminism and retrocausal influences.   It follows that all three of the main approaches for providing a causal explanation of Bell correlations, superluminal causes, superdeterminism and retrocausal influences, are unsatisfactory, and they are all unsatisfactory for the same reason. 

Our approach demonstrates that Bell's theorem can be understood as a statement about the possibility of a {\em causal} account of quantum correlations. This characterization is an improvement over the standard one for several reasons.   First, we believe that the question of what constitutes a {\em causal explanation} of correlations is more clearly defined than the question of what constitutes a {\em realist explanation} of those correlations.  Of course, if one likes,  one can take the notion of causal explanation to be an elucidation of the notion of realism at play in Bell's theorem.  In other words, one could take the view that an explanation should be described as {\em realist} only if it is causal. Indeed, the views of many proponents of {\em anti-realism} in quantum theory are aptly characterized as a denial of the need to provide a causal explanation of quantum correlations.  The second advantage of our characterization is that the fine-tuning criticism applies to {\em all} of the various attempts to provide a causal explanation of Bell inequality violations.  Accounts in terms of superluminal causes, superdeterminism or retrocausal influences are found to fall under a common umbrella.  The conspiratorial flavour of each such account can be formalized as a need for fine-tuning. 

\section{Causal structures and causal models}
\label{sec:causalmod}

The modern approach to the formal study of causality considers in some detail the significance of interventions and counterfactuals for defining the notion of a causal relation~\cite{Pearl2009,Spirtes2001}.  There is a large literature on whether these sorts of definitions are adequate~\cite{OxfordHandbook}.   Although questions of this sort are relevant to a discussion of Bell's theorem, they will not be the focus of this article.
We begin by describing the mathematical formalism that is relevant for describing the causal discovery algorithms in Refs.~\cite{Pearl2009} and~~\cite{Spirtes2001}.  We follow the presentation of these authors.

A \textbf{causal structure} is a set of variables $\mathbf{V}$ and a set of
ordered pairs of distinct variables $\left\langle X,Y\right\rangle $
specifying that $X$ is a direct cause of $Y$ relative to $\mathbf{V}.$

Being in a relationship of direct causation is a property that is defined
relative to the set of variables being considered. If one considers a
larger set which includes more variables, then what was a direct causal
relation in the first set might become a mediated causal relation in the
second.

Such causal structures can be represented conveniently by \textbf{directed
acyclic graphs} (DAGs).  A directed graph $G$ corresponds to a set of
vertices and a set of directed edges among the vertices (a vertex cannot be
connected to itself).  The acyclic property asserts that there are no directed paths in
the graph that begin and end at the same vertex.  DAGs represent causal
structures in the obvious manner: every variable in $\mathbf{V}$ is
represented by a vertex, and for every pair of variables $\left\langle
X,Y\right\rangle $ where $X$ is a direct cause of $Y,$ there is a directed
edge in the graph between the associated vertices~\footnote{One can imagine more general notions of causation wherein directed cycles are allowed, but we will not consider such notions here.}.

As is standard, we use the terminology of family relations in the obvious manner: if $X$
is a cause of $Y$, direct or mediated, then $X$ is said to be an \emph{ancestor} of $Y$, and $Y$ is said to be a \emph{descendent} of $X$. If $X$ is a direct cause of $Y$, then $X$ is said to be a \emph{parent} of $Y$.    The set of all parents of a variable $X$ will be denoted $Pa(X)$ and the set of all nondescendents of a variable $X$ will be denoted $Nd(X)$.  \color{black} The variables in the causal structure that have no parents will be called \emph{exogenous}, while those with parents will be called \emph{endogenous}. 

A \textbf{deterministic causal model} consists of a causal structure and a set $\Theta$ of
causal and statistical parameters.  The causal parameters describe the functional relations that fix the
values of every variable $X$ given its parents $Pa(X)$ in the causal structure, that is, for every $X$ they describe a function $f$ specifying $X=f(Pa(X))$.  The statistical parameters specify a probability distribution over the exogenous variables, that is, a distribution $P(X)$ for every exogenous $X$.

\begin{figure}[htbp]
	\begin{subfigure}[b]{0.2\textwidth}
		\centering
  		\includegraphics[width=\textwidth]{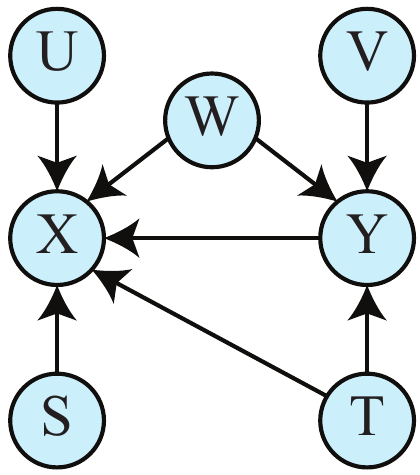}
	\end{subfigure}
	\quad
	\begin{subfigure}[b]{0.2\textwidth}
		\textbf{Model parameters}
		\begin{eqnarray*}
			&&P(S)\\
			&&P(T)\\
			&&P(U)\\
			&&P(V)\\
			&&P(W)\\
			&&X= f_X(S,T,Y,U,W)\\
			&&Y= f_Y(T,V,W)
		\end{eqnarray*}
	\end{subfigure}
\caption{An example of a deterministic causal model.}
\label{fig:detSTAB-graph}
\end{figure}

An example of a deterministic causal model is given in Fig.~\ref{fig:detSTAB-graph}.

The notion of a general causal model (not necessarily deterministic) can be explained as follows. We start with a deterministic causal model and modify it in a particular way.  When an exogenous variable $U$ is the parent of only a \emph{single} other variable, say $X$ (i.e. $U$ is not a common cause of two or more variables), it is possible to eliminate $U$ from the causal structure, and to replace the deterministic dependence of $X$ on its original set of parents with a probabilistic dependence on its new set of parents.  Specifically, if the deterministic causal model specifies that $X = f(Pa(X))$ for some function $f$ (here $Pa(X)$ includes $U$) then the new causal model specifies a conditional probability $P(X|Pa'(X))$ (here $Pa'(X)$ are the parents relative to the new causal structure, which excludes $U$). Specifically, the conditional probability is defined by $P(X|Pa(X))=\sum_U \delta_{X,f(Pa'(X),U)}P(U)$.

It follows that a general \textbf{causal model} consists of a causal structure and a set $\Theta$ of
causal-statistical parameters.  The causal-statistical parameters specify a conditional probability distribution for every variable given its causal parents, $P(X|Pa(X))$\footnote{ We have chosen to call the parameters of a general causal model ``causal-statistical'' because if the causal model arises from an underlying deterministic causal model,
then the conditional probabilities in the causal model fold together two different sorts of parameters from the underlying deterministic causal model: functional dependences of variables on their parents, which are causal parameters, and distributions over the local noise variables, which are statistical parameters. \color{black} 
\color{black}}.   Exogenous variables have the null set for their causal parents, so that to condition on their parents is not to condition at all.  Consequently, for the exogenous variables, the causal-statistical parameters simply specify the unconditioned distributions over each of these\footnote{Such models are sometimes called \emph{Markovian}. A more general sort of model, which allows bi-directed edges representing the existence of an unobserved common cause for a pair of variables, are called \emph{semi-Markovian}}.

\begin{figure}[htbp]
	\begin{subfigure}[b]{0.2\textwidth}
		\centering
  		\includegraphics[width=\textwidth]{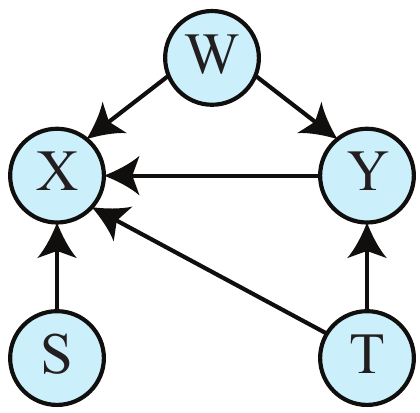}
	\end{subfigure}
	\quad
	\begin{subfigure}[b]{0.2\textwidth}
		\textbf{Model parameters}
		\begin{eqnarray*}
			&&P(S)\\
			&&P(T)\\
			&&P(W)\\
			&&P(X|S,T,Y,W)\\
			&&P(Y|T,W)
		\end{eqnarray*}
	\end{subfigure}
\caption{An example of a causal model consisting of a causal structure, represented by a directed acylic graph, and a set of causal-statistical parameters, specifying the probability of each variable conditioned on its parents.}
\label{fig:STAB-graph}
\end{figure}

An example of a general causal model is given in Fig.~\ref{fig:STAB-graph}.  It can be obtained from the deterministic causal model of Fig.~\ref{fig:detSTAB-graph} by eliminating the exogenous variables $U$ and $V$. (Note that one need not eliminate \emph{all} exogenous variables from a deterministic causal model to obtain a nondeterministic causal model-- for instance, $S$ and $T$ have not been eliminated in our example.)

Deterministic causal models are clearly a special case of causal models where all conditional probabilities correspond to deterministic functions. It is also clear that for any given causal model, one can always view it as arising from a deterministic causal model where some exogenous variables have been excluded.  To obtain such a deterministic extension of a causal model, it suffices to add new exogenous variables as parents of every endogenous variable in the model. For the rest of the article, we will focus on the general notion of a causal model, rather than on deterministic causal models.  

We pause to discuss briefly the possible interpretation of the probabilities in a causal model.
One could take a Bayesian attitude towards these probabilities.  In this case, the probability distribution on an exogenous variable $U$ represents an agent's degrees of belief about $U$, and the conditional probability $P(X|Pa(X))$ represents degrees of belief about $X$ given its parents.  Another possibility is to take a frequentist attitude towards the probabilities. This is arguably the position adopted by Pearl, who describes the auxiliary variables appearing in a deterministic extension of a causal model as `unmeasurable conditions that Nature governs by some undisclosed probability function' (\cite{Pearl2009},~p.~44).  One could even interpret the probabilities as \emph{propensities}, indicating an irreducible randomness in one's theory (an option that might be appealing to some when considering the possibility of explaining quantum correlations in terms of causal models).  Our conclusions here will be independent of this choice \footnote{Although we ultimately favor the Bayesian interpretation.}.

It is worth noting that the fact that exogenous variables are assumed to be independently distributed, which is part of the definition of a causal model, is a consequence of Reichenbach's principle.
The principle asserts that one must explain
all correlations by a causal mechanism, so that if two variables are correlated then either one is a cause of the other, or there is a common cause acting on both (this is not an exclusive or -- it could be that two variables have be related by both a common cause and a cause-effect relation).
In other words, the exogenous variables are by definition the variables that one takes to be uncorrelated. 

Consider the following question: given a causal model, what sorts of
correlations can be observed among the variables?  Clearly, there is a set of
joint distributions that are possible, depending on the causal-statistical parameters that we
add to the causal structure to get a causal model.

Consider the example from Fig.~\ref{fig:STAB-graph}.  The causal
model predicts that the joint distribution over all the variables should be
\begin{eqnarray}
P(X,Y,S,T,W)&=& P\left( W \right) P\left( S\right) P(T)\,P(Y|T,W)\,P(X|Y,S,T,W). \label{eq:exdist}
\end{eqnarray}
 To see this, it suffices to note that in the deterministic extension of this model, depicted in Fig.~\ref{fig:detSTAB-graph}, we have
\begin{eqnarray}
P(X,Y,S,T,W,U,V)&=& P\left(U\right) P\left( V \right) P\left( W \right) P\left( S\right) P(T)\, \delta_{Y,f_Y(T,V,W)} \, \delta_{X,f_X(S,T,Y,U,W)}, \label{eq:exdist}
\end{eqnarray}
where $\delta$ denotes the Kronecker delta function, $\delta_{X,Y}=1$ if and only if $X=Y$, and consequently,
\begin{eqnarray}
P(X,Y,S,T,W)&=& \sum_{U,V} P\left(U\right) P\left( V \right) P\left( W \right) P\left( S\right) P(T)\, \delta_{Y,f_Y(T,V,W)} \, \delta_{X,f_X(S,T,Y,U,W)}, \nonumber\\
&=&P\left( W \right) P\left( S\right) P(T)\,P(Y|T,W)\,P(X|Y,S,T,W)\,
\label{eq:exdist}
\end{eqnarray}
where $P(Y|T,W)\equiv \sum_V \delta_{Y,f_Y(T,V,W)} P\left(V\right) $ and $P(X|Y,S,T,W)\equiv \sum_U \delta_{X,f_X(S,T,Y,U,W)} P(U)$.
\color{black}

In general, a causal model with variables $\mathbf{V}\equiv \{ X_1,\dots,X_n\}$ predicts a joint distribution of the form
\begin{equation} \label{eq:jointdistribution}
P(X_1,\dots,X_n) = \prod_{i=1,\dots,n} P(X_i|\text{Pa}(X_i)).
\end{equation}
Essentially, one multiplies together the conditional probabilities for every
variable given its parents, all of which are specified by the causal model.
For a DAG that is not a complete graph (i.e not every pair of nodes is connected by an edge), the probability distributions that it supports are a subset of the possible distributions over those variables.

We now turn to another question: what properties do all distributions consistent with a given causal structure have in common?  In other words, what are the features of the joint probability distribution that depend \emph{only} on the causal structure and not on the causal-statistical parameters?  Conditional independence (CI) relations are an example of such properties, and they are the ones that most causal discovery algorithms to date have focussed upon.

Recall that variables $X$ and $Y$ are
conditionally independent given $Z,$ denoted
\begin{equation*}
(X\indep Y\text{ }|\text{ }Z)
\end{equation*}%
if any of the following three equivalent conditions hold 
\begin{eqnarray*}
1.\text{ }P(X|Y,Z) &=&P(X|Z)\qquad\forall y,z:P(Y=y,Z=z)>0, \\
2.\text{ }P(Y|X,Z) &=&P(Y|Z)\qquad\forall x,z:P(X=x,Z=z)>0, \\
3.\text{ }P(X,Y|Z) &=&P(X|Z)P(Y|Z)\qquad\forall z: P(Z=z)>0.
\end{eqnarray*}%
An intuitive account of each of these conditions is as follows: In the
context of already knowing $Z,$ (1) learning $Y$ teaches you nothing about $%
X $ (i.e. $Y$ teaches you nothing more about $X$ than what you already could
infer from knowing $Z$), (2) learning $X$ teaches you nothing about $Y$, and
(3) $X$ and $Y$ are independently distributed.  Note that \emph{marginal independence} of $X$ and $Y$, where $P(X,Y)=P(X)P(Y)$, is simply conditional independence where the conditioning set is the null set.

The definition of conditional independence implies that certain logical inferences hold among CI relations. In other words, in a complete set of CI relations, the CI relations need not be logically independent of one another.  In particular, the
\emph{semi-graphoid axioms} specify some inferences that can be drawn among CI
relations. They are:
\begin{eqnarray*}
\mbox{Symmetry: } 		&&	(X\indep Y\, |\,Z) 		\Leftrightarrow 		(Y\indep X\, |\,Z)\\
\mbox{Decomposition: } 	&& 	(X\indep YW\,|\,Z) 	\Rightarrow 		(X \indep Y |\,Z) \\
\mbox{Weak Union: } 	&&	(X\indep YW\,|\,Z)	\Rightarrow 		(X \indep Y \,| \, ZW) \\
\mbox{Contraction: } 	&&	(X \indep Y\,|\,Z) 	\mbox{ and }		 	(X\indep W\, |\, ZY)	\Rightarrow (X\indep YW \,|\,Z)
\end{eqnarray*}
Any set of variables can be considered as a new variable, so each of the variables $X$,$Y$, $W$ and $Z$ appearing in the axioms should be understood as possibly representing a set of variables.  These axioms are quite intuitive.  Decomposition, for instance, states that if, in the context of knowing $X$,
learning $W$ and $Y$ teaches you nothing about $U,$ then learning $W$ \emph{
alone }teaches you nothing about $U$.

Note that if one wants to specify \emph{all} the CI relations that hold for a given
probability distribution, it suffices to specify a \emph{generating set},
defined to be a set from which the rest can be obtained by the semi-graphoid axioms.  In this
paper, the conditional independence relations will typically be specified by
a generating set.

With these tools in hand, we can now discuss the central result concerning what properties of a joint probability distribution can be inferred from the causal structure.

\begin{theorem}[Causal Markov condition] In the joint distribution induced by
a causal structure, every variable $X$ is conditionally
independent of its nondescendants given its parents,
\begin{equation*}
(X\indep \text{\textrm{Nd}}\left( X\right) |\text{ \textrm{Pa}}\left( X\right) ).
\end{equation*}
\end{theorem}

This result follows from Eq.~\eqref{eq:jointdistribution} because
\begin{align}
P(X|\text{Pa}(X),\text{Nd}(X))
&= \frac{P(X,\text{Pa}(X),\text{Nd}(X))}{P(\text{Pa}(X),\text{Nd}(X))}, \notag  \\
&=\frac{P(X|\text{Pa}(X))\prod_{Y\in \text{Pa}(X),\text{Nd}(X)} P(Y|\text{Pa}(Y)) }
{\prod_{Y\in \text{Pa}(X),\text{Nd}(X)} P(Y|\text{Pa}(Y)) }, \notag  \\
&=P(X|\text{Pa}(X)).
\end{align}

The causal Markov condition implies a CI relation for every variable that is not exogenous in the causal structure. One can then infer additional CI relations from these by the semi-graphoid axioms.

To see these ideas in action, consider again the example from Fig.~\ref{fig:STAB-graph}.  It turns out that $(Y \indep S | T)$ for this causal structure, as we now demonstrate. Applying the causal Markov condition to $Y$, one infers that $(Y \indep XS | WT)$.  Applying it to $W$, $S$ and $T$ one infers $(W\indep ST)$, $(S\indep WT)$ and $(T \indep WS)$ respectively.  By the decomposition axiom, $(Y \indep XS | WT)$ implies $(Y \indep S | WT)$.  From the contraction axiom, $(Y \indep S | WT)$ and $(S \indep WT)$ imply $(S\indep YWT)$.  Finally, from weak union we obtain $(S\indep YW|T)$ and then from decomposition again we have $(S\indep Y|T)$, which is equivalent by symmetry to $(Y\indep S|T)$.

We see that it can be rather laborious to infer CI relations from the causal Markov condition and the semi-graphoid axioms.  Fortunately, there is a graphical criterion for identifying such relations, known as \emph{d-separation} \cite{Pearl2009}. We will not dwell on this notion here, but we present a brief introduction in App.~\ref{app:dsep}.

Note that in addition to the CI relations that are implied by the causal structure, there may be additional CI relations that are implied by the particular values of the causal-statistical parameters.   Such additional CI relations are problematic for causal discovery algorithms, as we shall see.

\section{Causal discovery algorithms}
\label{sec:causalalg}

We have described the correlations that are possible
for a given causal structure.  Causal discovery algorithms seek to solve
the inverse problem: starting from correlations among observed variables,
can one infer which causal structures might account for these correlations?
Researchers in this area have indeed devised some schemes for narrowing down
the set of causal structures that can yield a natural explanation of the
correlations, wherein the notion of naturalness at play is one that we shall
make explicit shortly.  The algorithms look to the conditional
independences among the variables to infer information about the causal structure.

In general, causal discovery algorithms may be applied directly to experimental data and in this case one needs to deal with the subtle issue of how to infer conditional independence relations from a finite sample of a probability distribution.  However, in what follows we are going to apply the causal discovery algorithms directly to the distributions prescribed by quantum theory, so we needn't worry about this subtlety.

\begin{figure}[h]
        \begin{subfigure}[b]{0.2\textwidth}
                	\centering
                	\includegraphics[width=\textwidth]{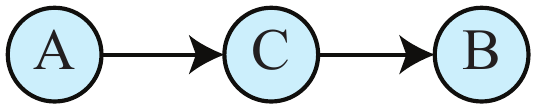}
		\caption{}
                	\label{fig:markovchain1}
        \end{subfigure}
        \quad\quad
        \begin{subfigure}[b]{0.2\textwidth}
                	\centering
                	\includegraphics[width=\textwidth]{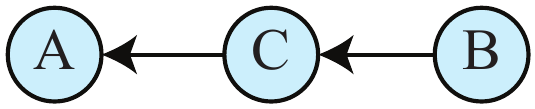}
		\caption{}
                	\label{fig:markovchain2}
        \end{subfigure}
        \quad
        \begin{subfigure}[b]{0.2\textwidth}
                	\centering
                	\includegraphics[width=\textwidth]{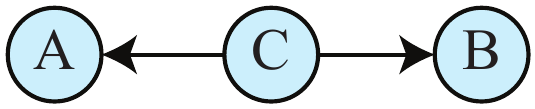}
		\caption{}
                	\label{fig:markovchain3}
        \end{subfigure}
 \caption{The three causal models consistant with the CI relation $(A\indep B\,|\,C)$}
        \label{fig:markovchain}
\end{figure}

It is worth reviewing a few basic facts about the output of causal discovery algorithms. First of all, two different causal structures might support precisely the same probability distributions, so that observation of
one of these distributions necessarily leaves one ignorant about which
causal structure is at play. As an example, for three variables, the three
causal structures show in Fig.~\ref{fig:markovchain} all support the same set of probability distributions -- those wherein $A$ and $B$ are conditionally independent given $C$ (these are the DAGs wherein $A$ and $B$ are d-separated given $C$). (The general conditions under which two causal structures are observationally equivalent is given by theorem 1.2.8 in Ref.~\cite{Pearl2009}.)

It follows that causal discovery algorithms will necessarily sometimes yield an equivalence class of
causal structures.  When this occurs, additional information is required if one is to narrow down the causal structure to a unique possibility, for instance information about the temporal order of some of the variables.

\begin{figure}[h]
	\centering
	\includegraphics[width=0.2\textwidth]{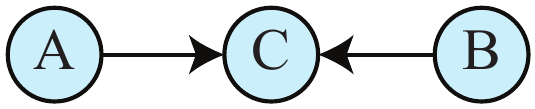}
 \caption{The natural causal model for the set of CI given in Eq.~(\ref{eqn:a-ind-b}).}
       \label{fig:a-ind-b}
\end{figure}

Despite this, one can often narrow down the field of causal possibilities
significantly.  To get a feeling for how this works, it is useful to start
with a very simple example.  Suppose that one has three binary-valued
variables, denoted $A,$ $B$ and $C.$ Suppose further that the joint
distribution over the triple, $P(A,B,C)$ is such that%
\begin{eqnarray}
(A \indep B) \;\;\;\textrm{i.e.}\;\;\; P(A,B) &=&P(A)P(B),  \notag \\
(A \not\indep C) \;\;\;\textrm{i.e.}\;\;\; P(A,C) &\neq &P(A)P(C),  \notag \\
(B \not\indep C) \;\;\;\textrm{i.e.}\;\;\; P(B,C) &\neq &P(B)P(C).
\label{eqn:a-ind-b}
\end{eqnarray}%
What is the \emph{natural} causal explanation for this sort of correlation?
It is as shown in Fig.~\ref{fig:a-ind-b}. The marginal independence of $A$ and $B$ is explained by their being causally independent.

\begin{figure}[h]
	\centering
	\includegraphics[width=0.2\textwidth]{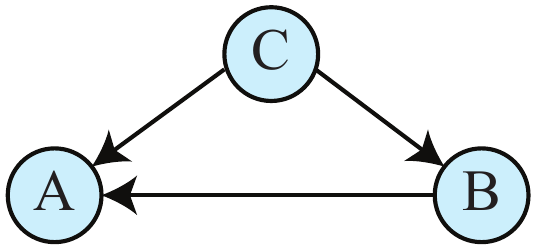}
 \caption{An unnatural causal model for the set of CI given in Eq.~(\ref{eqn:a-ind-b})}
       \label{fig:a-ind-b-alt}
\end{figure}

However, there are other possible causal explanations, such as the one given in Fig.~\ref{fig:a-ind-b-alt}. The reason this is a possible explanation is because there are two causal
mechanisms by which $A$ and $B$ could become correlated, and it \emph{
could be }that the two types of correlations combine in such a way as to
leave $A$ and $B$ marginally independent.
For this to happen, however, the parameters in the causal model cannot be chosen arbitrarily and it is in this sense that the explanation is less natural than the one provided by Fig.~\ref{fig:a-ind-b}.

An example helps to make all of this more explicit.  
We adopt the following notational convention (inspired by the representation
of mixtures in quantum theory)
\begin{eqnarray*}
P(A) &=&[x] \text{ means }P\left( A=x\right) =1, \\
P(A,B) &=&[x][y]\equiv \lbrack xy]\text{ means } P\left( A=x,B=y\right) =1.
\end{eqnarray*}%
\ Consider the following joint distribution, which has the dependences
described in Eq.~(\ref{eqn:a-ind-b}),
\begin{equation}
P(A,B,C)=\frac{1}{4}[000]+\frac{1}{4}[010]+\frac{1}{4}[100]+\frac{1}{4}[111].
\label{PABC}
\end{equation}
We can easily verify that
\begin{equation*}
P\left( A,B\right) =\left( \frac{1}{2}[0]+\frac{1}{2}[1]\right) \left( \frac{%
1}{2}[0]+\frac{1}{2}[1]\right) ,
\end{equation*}%
so that $A$ and $B$ are indeed marginally independent.  We also have
\begin{equation*}
P(A,C)=P(B,C)=\frac{1}{2}[00]+\frac{1}{4}[10]+\frac{1}{4}[11],
\end{equation*}%
so that $A$ and $C$ are marginally dependent, as are $B$ and $C$ .

The natural explanation is achieved by assuming that the causal structure is as given in Fig.~\ref{fig:a-ind-b}, and the priors over the exogenous variables and the conditional
probabilities for the endogenous variables are as follows:%
\begin{eqnarray*}
P(A) &=&\frac{1}{2}[0]+\frac{1}{2}[1], \\
P(B) &=&\frac{1}{2}[0]+\frac{1}{2}[1], \\
P(C|A,B) &=&\left[ A\cdot B\right] ,
\end{eqnarray*}%
where $A\cdot B$ denotes the product of the values of $A$ and $B.$ Thus in this
causal model, $A$ and $B$ are each chosen uniformly at random, and $C$ is
obtained as their product (equivalently, as the logical AND of $A$ and $B$).
 One can easily verify that $P(A)P(B)P(C|A,B)$ yields the distribution of
Eq. (\ref{PABC}).

The alternative explanation assumes the causal structure of Fig.~\ref{fig:a-ind-b-alt}, with
parameters
\begin{eqnarray*}
P(C) &=&\frac{3}{4}[0]+\frac{1}{4}[1], \\
P(B|C &=&0)=\frac{2}{3}[0]+\frac{1}{3}[1], \\
P(B|C &=&1)=[1], \\
P(A|B &=&0,C=0)=\frac{1}{2}[0]+\frac{1}{2}[1], \\
P(A|B &=&1,C)=[C].
\end{eqnarray*}%
(We need not specify $P(A| B=0, C=1)$ because $P(B=0, C=1)=0$.)  The joint distribution one obtains is again that of Eq. (\ref{PABC}).

The difference between the two explanations becomes clear when we vary the
parameters.  If we change the parameters in the first model, for instance
to
\begin{eqnarray*}
P(A) &=&w[0]+(1-w)[1], \\
P(B) &=&w^{\prime }[0]+\left( 1-w^{\prime }\right) [1], \\
P(C|A,B) &=&w^{\prime \prime }\left[ AB\right] +(1-w^{\prime \prime })[A\oplus B],
\end{eqnarray*}
where $\oplus$ denotes addition modulo 2, then the joint distribution is no longer of the form of Eq.~(\ref{PABC}),
but it is still true that $A$ is independent of $B,$ while $A$ and $C$ are
dependent, and $B$ and $C$ are dependent. On the other hand, modifications
to the parameters in the second model do not preserve the pattern of
dependences and independences among $A,$ $B$ and $C.$

The first causal structure explains the pattern of statistical dependences
and independences in a manner that is robust to changes in the parameters of
the causal model, whereas the second causal structure does not. Causal
discovery algorithms therefore favour the first model over the second.

In the example we have used, all of the variables in the causal model were observed variables.  In general (and especially in a quantum context), one might only observe a subset of the variables that are part of the causal model. Even in this case, however, one should prefer those causal models wherein the conditional independences in the probability distribution over the observed variables are stable to changes in the causal-statistical parameters.

This is the main assumption of the causal discovery algorithms, usually
called \emph{faithfulness }\cite{Spirtes2001} or \emph{stability}\cite{Pearl2009}.   For a physicist, it is natural to call this an assumption of \emph{no fine-tuning}. \color{black}  It is the key
assumption in our analysis, so we highlight it:

\begin{quote}
\textbf{Faithfulness (no fine-tuning): }The probability distribution induced by a causal
model M (over the variables in M or some subset thereof) is faithful (not fine-tuned) if its conditional independences continue to hold for any variation of the causal-statistical parameters in M.
\end{quote}

In other words, all conditional independences should be a consequence of the causal structure alone, not a result of the causal-statistical parameters taking some particular set of values.
If one assumes a uniform prior over the space of causal-statistical parameters, then the parameter choices that can explain 
conditional independence relations that are not implied by the causal structure are found to have measure zero.

The second major assumption of CI-based causal discovery algorithms is an appeal to Occam's razor, an assumption that one should favour the most simple or most minimal model that explains the statistics.  Again, it can be applied both for the case where the observed variables are all the variables in the causal model, or the case where they are some subset thereof.

A causal model $M$ will be said to \emph{simulate} another causal model $M'$ on a set of variables $V$ if for every choice of causal-statistical parameters on $M'$, there is a choice of causal-statistical parameters on $M$ such that $M$ yields the same distribution over $V$ as $M'$ does. We can now define the assumption of minimality.

\begin{quote}
\textbf{Minimality: } Given two causal models $M$ and $M'$ that induce a given probability distribution over a set of observed variables $V_O$ (in general a subset of the variables postulated by each causal model), if $M'$ can simulate $M$ on $V_O$ but $M$ cannot simulate $M'$ on $V_O$, then $M$ is preferred to $M'$ as a causal explanation of the probability distribution over $V_O$.
\end{quote}

At first sight, it might seem odd to prefer $M$ over $M'$ given that $M$ is consistent with fewer distributions over $V$ than $M'$ is.  But the fact that $M$ can explain \emph{less} than $M'$ implies that $M$ is \emph{more falsifiable} than $M'$, and in the version of Occam's razor espoused by CI-based causal discovery algorithms, the degree of falsifiability is the figure of merit that one seeks to optimize. More falsifiable theories are to be preferred because, in Pearl's words, ``they provide the scientist with less opportunities to overfit the data `hindsightedly' and therefore command greater credibility if a fit is found'' (\cite{Pearl2009},~p.~49).  It follows that a causal model is deemed most simple if it has the \emph{least} expressive power, while still doing justice to the observed probability distribution.  Note that $M$ might be preferred to $M'$ as a causal explanation of the probability distribution over $V_O$ even though $M$ may require \emph{more} latent variables and/or \emph{more} causal arrows than $M'$;  ``the preference for simplicity [...] is gauged by the expressive power of a structure, not by its syntactic description.'' (\cite{Pearl2009},~p.~46).
We will see some examples of the consequences of the assumption of minimality shortly.

It is worth remembering that causal discovery algorithms are fallible.
They are best considered a heuristic, an inference to the best explanation.
Indeed, Pearl likens the faithfulness assumption in causal discovery to the
following kind of inference: you see a chair before you and infer that there
is a single chair rather than two chairs positioned such that the one hides
the other (\cite{Pearl2009},~p.~48).  The task of causal discovery can be understood
as ``an inductive game that scientists play against Nature'' (\cite{Pearl2009},~p.~42).

\subsection{Example of causal discovery assuming no latent variables}
\label{sec:causal-nohvar}

Variables that are not observed but which are causally relevant are called \emph{latent} variables, or \emph{hidden} variables. In this section, we assume
that the observed variables are the \emph{only} causally relevant
variables, i.e. that there are no hidden
variables.  We look at a particular example of how faithfulness can help to determine
candidate causal structures from a pattern of dependences in this case.  The scheme is equivalent to the one introduced by Wermuth and Lauritzen \cite{Wermuth1983}.

Suppose one is interested in answering the question ``Does smoking cause lung cancer?''
 For each member of a population of individuals, the value of a
variable $S$ is known, indicating whether the individual smoked or not, and the value
of a variable $C$ is known, indicating whether they developed cancer or not.
Suppose a correlation between $S$ and $C$ is observed. Furthermore, suppose
that one also has access to a third variable $T,$ indicating whether the individual
had tar in their lungs or not, and suppose that it is found that $S$ and $C$ are conditionally
independent given $T.$ In other words, after conditioning on whether or
not there is tar in the lungs, smoking and lung cancer are no longer
correlated. Finally, imagine that these three variables are assumed to be
the only causally relevant ones (we will consider the alternative to this
assumption further on). What causal structure is natural given the observed
conditional independence relation?
Because we wish to make it very clear how these algorithms work, we will not
simply specify what causal structure they return. Instead, we will look
``under the hood'' of these algorithms.

We begin by considering every possible hypothesis about the causal ordering. A
causal ordering of variables is an ordering wherein causal influences can
only propagate from one variable to another if the second is higher in the
order than the first.

\begin{figure}[h]
	\centering
	\includegraphics[width=0.2\textwidth]{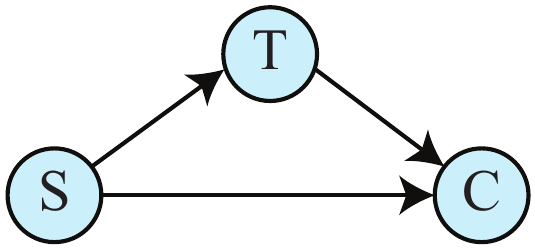}
 \caption{The most general DAG for the causal ordering $S<T<C$.}
       \label{fig:smoke-STC-gen}
\end{figure}

\begin{figure}[h]
	\centering
	\includegraphics[width=0.2\textwidth]{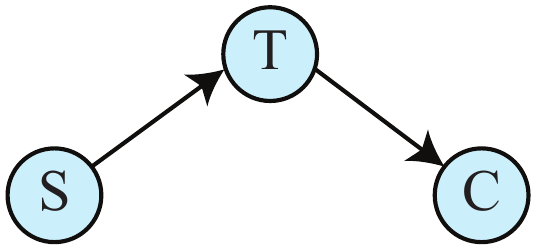}
 \caption{DAG that captures $(S\indep C\,|\,T)$ for the causal ordering $S<T<C$.}
       \label{fig:smoke-STC-ind}
\end{figure}

For instance, consider the causal ordering $S<T<C.$ The most general
causal structure consistent with such an ordering is given in Fig.~\ref{fig:smoke-STC-gen}.
To get a causal model, we need to supplement this with conditional
probabilities of every variable given its parents, that is, $P(S),P(T|S),$
and $P(C|T,S).$ The joint distribution that this model defines is simply%
\begin{equation*}
P(S,T,C)=P(S)P(T\,|\,S)P(C\,|\,T,S).
\end{equation*}%
Given that \emph{any} distribution can be decomposed in this form, by
choosing the conditional probabilities appropriately, we can model \emph{any
}joint distribution $P(S,T,C).$ But now we make use of the additional
information we have about the joint distribution, namely that $(S\indep C\,|\,T).$
This implies that we can take the parameters in
the causal model to be such that $P(C\,|\,T,S)=P(C\,|\,T)$, so that the joint
distribution can be written as
\begin{equation*}
P(S,T,C)=P(S)P(T\,|\,S)P(C\,|\,T),
\end{equation*}%
and we can drop the causal arrow from $S$ to $C,$ so that the underlying causal structure is simply given by Fig.~\ref{fig:smoke-STC-ind}.

This simplified causal structure cannot generate an arbitrary probability
distribution, but it \emph{can }generate one wherein $(S\indep C\,|\,T).$
It is a candidate for the true causal structure.

\begin{figure}[h]
	\centering
	\includegraphics[width=0.2\textwidth]{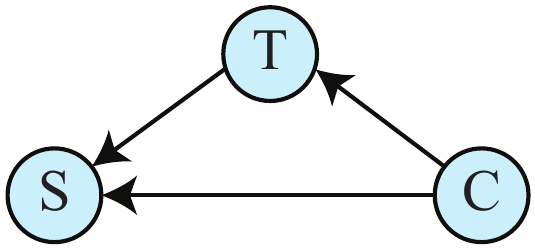}
 \caption{The most general DAG for the causal ordering $C<T<S$.}
       \label{fig:smoke-CTS-gen}
\end{figure}

\begin{figure}[h]
	\centering
	\includegraphics[width=0.2\textwidth]{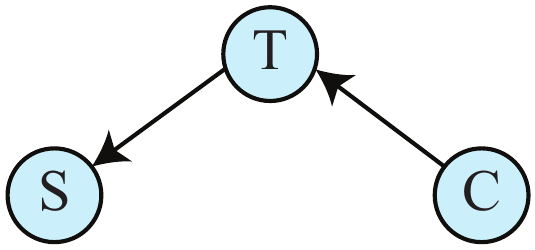}
 \caption{DAG that captures $(S\indep C\,|\, T)$ for the causal ordering $C<T<S$.}
       \label{fig:smoke-CTS-ind}
\end{figure}

One then simply repeats this procedure for every possible choice of the
causal ordering. For instance, for the ordering $C<T<S,$ the most general
causal structure is the one shown in Fig.~\ref{fig:smoke-CTS-gen}.
The decomposition of the joint probability corresponding to this causal structure is%
\begin{equation*}
P(S,T,C)=P(C)P(T|C)P(S|C,T),
\end{equation*}%
but the constraint $(S\indep C|T)$ implies that one substitute $P(S\,|\,C,T)=P(S\,|\,T)$
in the causal model.  Therefore, by the assumption of minimality, we drop the causal arrow
from $C$ to $S$, yielding 
a causal structure of the form given in Fig.~\ref{fig:smoke-CTS-ind}. So this is another possible causal structure.

\begin{figure}[h]
	\centering
	\includegraphics[width=0.2\textwidth]{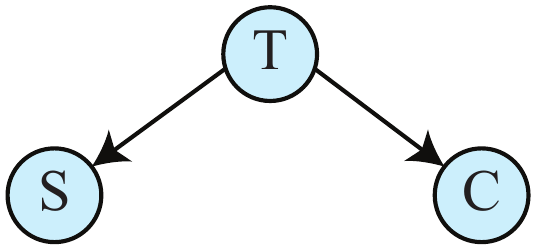}
 \caption{DAG that captures $(S\indep C\,|\, T)$ for the causal orderings $T<S<C$ and $T<C<S$..}
       \label{fig:smoke-TSC-ind}
\end{figure}

Sometimes different causal orderings lead to the same causal structure, for
instance, the orderings $T<S<C$ and $T<C<S$ both yield the structure given in Fig.~\ref{fig:smoke-TSC-ind}. %

\begin{figure}[h]
        \begin{subfigure}[b]{0.2\textwidth}
                	\centering
                	\includegraphics[width=\textwidth]{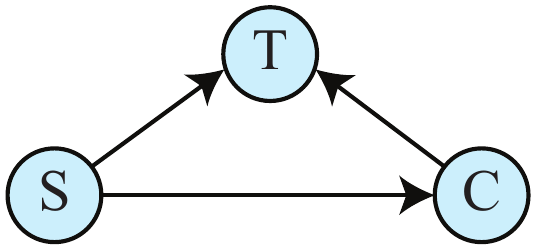}
		\caption{}
                	\label{fig:smoke-SCT}
	\end{subfigure}
        \quad
        \begin{subfigure}[b]{0.2\textwidth}
                	\centering
                	\includegraphics[width=\textwidth]{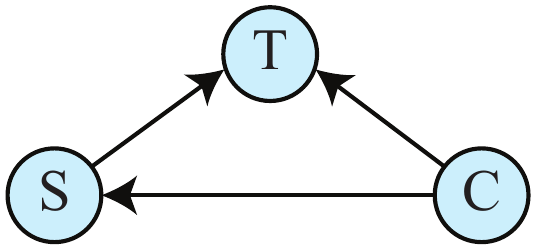}
            	\caption{}
                	\label{fig:smoke-CST}
        \end{subfigure}
        \caption{DAGs that capture $(S\indep C\,|\, T)$ for the causal orderings $S<C<T$ (\ref{fig:smoke-SCT}) and $C<S<T$ (\ref{fig:smoke-CST}).}
        \label{fig:smoke-SCT-CST}
\end{figure}

Other causal orderings, such as $S<C<T$ and $C<S<T$ are such that the
conditional independence constraint does not lead to any simplification of
the causal structure. \ For instance, for $S<C<T,$ the joint distribution
decomposes as $P(S,T,C)=P(S)P(C|S)P(T|C,S),$ and none of the terms on the
right-hand side can be simplified by $(S\indep C|T).$  These two orderings lead to the two causal structures in Fig.~\ref{fig:smoke-SCT-CST}.

Therefore, in this example, the six possible causal orderings have led to five candidates for the causal structure, depicted in Figs~\ref{fig:smoke-STC-ind}, \ref{fig:smoke-CTS-ind}, \ref{fig:smoke-TSC-ind} and~\ref{fig:smoke-SCT-CST}. \ However, the two causal structures shown in Fig.~\ref{fig:smoke-SCT-CST} do not satisfy stability, so only the other three are viable.

Suppose finally that in addition to the information about conditional
independence, one has information which rules out certain causal orderings.
For instance, in the example we are considering, suppose one has the
additional information that tar in the lungs always appears \emph{after} a
person has smoked, never before. It is then reasonable to rule out any
causal structure that has $T<S.$ This rules out Figs~\ref{fig:smoke-CTS-ind} and \ref{fig:smoke-TSC-ind}. \ At the end, the only candidate causal structure which is left is the one given in Fig.~\ref{fig:smoke-STC-ind}, which says that smoking causes tar in the lungs which causes lung cancer.

Of course, it needn't be the case that these observed variables are the only
ones that are causally relevant. For instance, there might be an unobserved genetic
factor that predisposes people both to smoke and to
develop lung cancer. Indeed, tobacco companies were quick to point out the possibility of
explaining the observed correlation between smoking and cancer in terms of
such a genetic factor. So it is useful also to have causal discovery
algorithms that allow for latent variables.

Before moving on to algorithms that posit latent variables, we pause to note
that the algorithm described here is proven to be correct in the sense
that if there exists a set of causal structures that are minimal and
faithful to the observed correlations, then the algorithm will return these structures~\cite{Wermuth1983}.

More efficient versions of this algorithm are described elsewhere, for
instance, the Inductive causation (IC) algorithm described in Pearl~\cite{Pearl2009}, which is equivalent to the SGS algorithm of Spirtes, Glymour and Scheines~\cite{Spirtes2001}. There have also been many proposals to further improve the efficiency of these algorithms (See Refs.~\cite{Pearl2009} and~\cite{Spirtes2001} for details) .  These algorithms have been proven to be correct in the sense that if there exist causal models that are minimal and faithful, then the algorithms will return them.

\subsection{Example of causal discovery allowing for latent variables}
\label{sec:causal-hvar}

Causal discovery in the case where one allows latent variables is more complicated.
We begin by considering some of the consequences of the assumption of minimality for causal models with latent variables.

First of all, it is clear that one needn't consider any causal models wherein a latent variable
\emph{mediates} a relation between two observed variables, because the set of
distributions over the observed variables that can be explained by such a model is no greater
than the set that can be explained by simply postulating a direct causal
influence between the observed variables.  Similarly, positing a latent variable that is a common \emph{effect} of the observed variables does not change the distributions that can be supported on the observed variables.  Latent variables have nontrivial consequences for the observed distribution only when they act as \emph{common causes} of the observed variables.

Consider the following suggestion for a causal explanation of
the correlations among a set of observed variables: there are no causal
influences among any of the observed variables, but there is a single latent
variable that has a causal influence on each of them. By choosing the
latent variable to take as many values as there are valuations of the
observed variables, one can explain \emph{any} correlation among the
observed variables in this way. However, if there exists another causal model that
can only reproduce a smaller set of possible correlations, while reproducing
the observed correlations, then the principle of minimality dictates that we
should prefer the latter. Of course, one could imagine that further
investigations (involving interventions, for instance) might vindicate the
explanation that is less falsifiable over the one that is more falsifiable.
This simply is another reminder that causal discovery algorithms are not
infallible --- they are heuristics for identifying the most plausible causal
explanations given the evidence.

Now we come to the most subtle part of the causal discovery algorithms that posit latent variables.
There is a difference between applying the criterion of minimality among a set of causal structures that are consistent with a given \emph{distribution} over the observed variables and applying the criterion of minimality among a set of causal structures that are consistent with a given set of \emph{conditional independence relations} over the observed variables.
As we've mentioned before, the algorithms described in Refs. \cite{Pearl2009} and \cite{Spirtes2001} look only at the CI relations and consequently they follow the latter course.  This choice is a significant shortcoming of many prominent causal discovery algorithms, but we will defer this criticism until the end of this section.

For the moment, we simply explain the consequences of this choice.  To do so, it is useful to divide the causal structures that are consistent with a given distribution over a set of observed variables into two sorts.  The first kind is such that all the latent variables it posits are common causes for at most two of the observed variables. We'll say that such a causal structure is limited to \emph{pairwise common causes}.  The other kind is unrestricted, so that more than two observed variables can be directly influenced by a single latent variable.

It is possible to show~\cite{Verma1993}  that for a given set of CI relations among a set of observed variables, if a causal model $M$ generates those CI relations faithfully (that is, as a consequence of the causal structure, rather than the causal-statistical parameters), then there is another causal model $M'$ that achieves the same CI relations faithfully but which is limited to pairwise common causes.  The assumption of minimality makes $M'$  preferred to $M$.

Therefore, if one is only applying the criterion of minimality among a set of causal structures that are consistent with the CI relations among the observed variables, then one need only look among causal models that incorporate pairwise common causes.  
This is precisely what the standard causal discovery algorithms do. 

\begin{figure}[h]
        \begin{subfigure}[b]{0.2\textwidth}
                	\centering
                	\includegraphics[width=\textwidth]{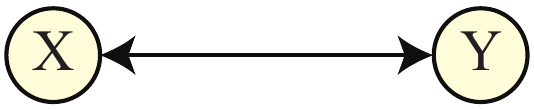}
                	\label{fig:XY-darrow}
        \end{subfigure}
                      \begin{subfigure}[b]{0.05\textwidth}
                	\centering
                	$\equiv$\vspace{2em}
        \end{subfigure}
        \begin{subfigure}[b]{0.2\textwidth}
                	\centering
                	\includegraphics[width=\textwidth]{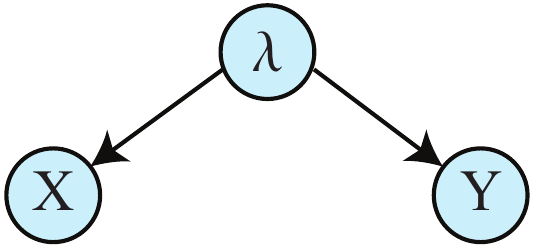}
                	\label{fig:XY-hvar}
        \end{subfigure}
        \caption{The interpretation of a bidirected edge in terms of a DAG.}
        \label{fig:XY-latentcause}
\end{figure}

\begin{figure}[h]
\begin{eqnarray*}
        \begin{subfigure}{0.2\textwidth}
                	\centering
                	\includegraphics[width=\textwidth]{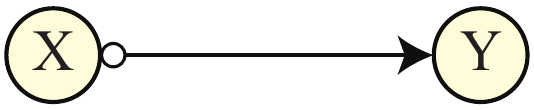}
        \end{subfigure}
                	&\equiv&
        \begin{subfigure}{0.2\textwidth}
                	\centering
                	\includegraphics[width=\textwidth]{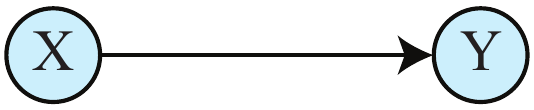}
	        \end{subfigure}
	     	\hspace{1em} or \hspace{1em}
      	 \begin{subfigure}{0.2\textwidth}
                	\includegraphics[width=\textwidth]{XY-hvar}
        	\end{subfigure}
		\hspace{1em} or \hspace{1em}
      	 \begin{subfigure}{0.2\textwidth}
                	\includegraphics[width=\textwidth]{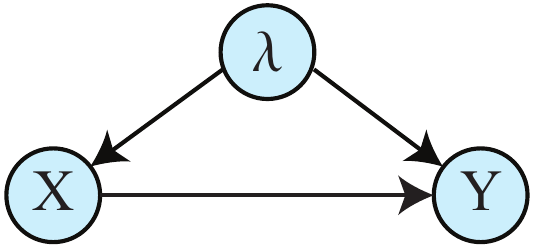}
        	\end{subfigure}		
       \end{eqnarray*}
        \caption{The interpretation of a directed edge with a circle at its tail in terms of DAGs.}
        \label{fig:XY-ambcause}
\end{figure}

\begin{figure}[h]
\begin{eqnarray*}
        \begin{subfigure}{0.2\textwidth}
           \centering
                	\includegraphics[width=\textwidth]{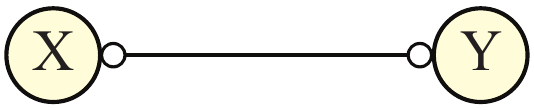}
        \end{subfigure}
                	&\equiv&
        \begin{subfigure}{0.2\textwidth}
                	\centering
                	\includegraphics[width=\textwidth]{XY-arrow}
	        \end{subfigure}
	        \hspace{1em} or \hspace{1em}
      	 \begin{subfigure}{0.2\textwidth}
                	\includegraphics[width=\textwidth]{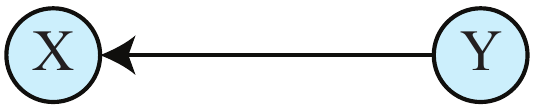}
        	\end{subfigure}
		\hspace{1em} or \hspace{1em}
      	 \begin{subfigure}{0.2\textwidth}
                	\includegraphics[width=\textwidth]{XY-hvar}
        	\end{subfigure}
		 \\&&or\hspace{1em}
      	 \begin{subfigure}{0.2\textwidth}
               	\vspace{1em}
                	\includegraphics[width=\textwidth]{XY-hvararrow}
        	\end{subfigure}	
		\hspace{1em} or \hspace{1em}	
      	 \begin{subfigure}{0.2\textwidth}
               	\vspace{1em}
                	\includegraphics[width=\textwidth]{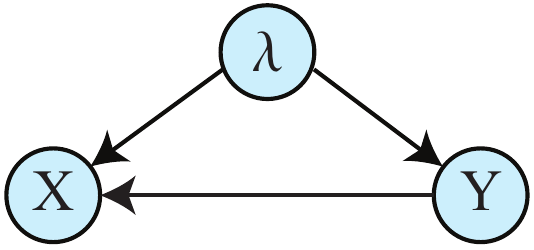}
        	\end{subfigure}		
       \end{eqnarray*}
        \caption{The interpretation of an undirected edge with circles at head and tail in terms of DAGs.}
        \label{fig:XY-equiv}
\end{figure}

There is a simplified graphical language for representing the set of causal structures that can be output by these algorithms.  Rather than using a DAG that includes both the latent and the observed variables in the causal structure, one uses a graph which only includes the observed variables as nodes but uses a larger variety of edges among these nodes to specify the causal relation that might hold among the associated variables.
For instance, a double-headed arrow between variables $X$ and $Y$ signifies that there is a common cause of $X$ and $Y$ (Fig.~\ref{fig:XY-latentcause}). An
arrow that has a circle rather than an arrowhead at one end represents either a common cause or a direct causal influence or both (Fig.~\ref{fig:XY-ambcause}).
Finally, an undirected edge with a circle at both its head and its tail represents any of the five possible ways in which a pair of variables might be related (Fig.~\ref{fig:XY-equiv}).  In this way, a set of causal structures that include latent variables can be summarized in a single graph.  Following Pearl, we call such graphs \emph{patterns}~\footnote{More precisely, the analogue of the particular graphs we consider here are Pearl's ``marked patterns''. These have also been called ``partially oriented inducing path graphs'' in SGS.  We will follow SGS's notational convention rather than Pearl's when drawing such graphs.}.

In order to infer the causal structures with only pairwise common causes that are consistent with a given pattern, it is \emph{not} sufficient to simply substitute for every undirected edge (or bi-directed edge or directed edge with decorated tail) all the possibilities consistent with that edge, as enumerated in Figs.~\ref{fig:XY-latentcause},~\ref{fig:XY-ambcause} and \ref{fig:XY-equiv}.  One must eliminate some of the combinations.  The definition of a \emph{v-structure} in a DAG is a head-to-head collision of two arrows on a node such that the parents do not exert any direct causal influence on one another. The prescription for finding all the DAGs consistent with a pattern is to consider all the combinations of possibilities that \emph{do not create a new v-structure}.

The IC* algorithm described in Pearl~\cite{Pearl2009} (which is equivalent to the Causal Inference (CI) algorithm described in SGS~\cite{Spirtes2001}) takes conditional independence relations as input and returns a pattern.  This algorithm is correct in the sense that if there exist causal structures with only pairwise common causes that are
faithful to the observed CI relations, then the algorithm will return the minimal structures within this set.\footnote{Note, however, that the existence of a causal structure that reproduces the CI relations does not guarantee the existence of one that reproduces the observed distribution, as we will see at the end of this section. In this sense, the algorithm may still fail to return a valid causal explanation of the observed distribution.}  We will not review the details of the algorithm here, but we will apply it to a simple example to get a feeling for how it works.

\begin{figure}[ht]
	\centering
	\includegraphics[width=0.2\textwidth]{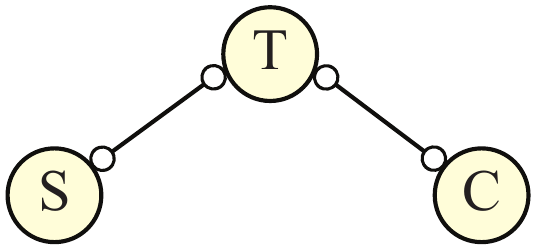}
 \caption{Output pattern of IC* algorithm for input $S \indep C|T$.}
       \label{fig:smokeICstar}
\end{figure}

Consider the smoking example again, where the observed variables $S,T$ and $C$ are found to satisfy $S\indep C\,|\,T$. The pattern returned by the IC* algorithm in this case is shown in Fig.~\ref{fig:smokeICstar}.

\begin{figure*}
	\centering
	\includegraphics[width=\textwidth]{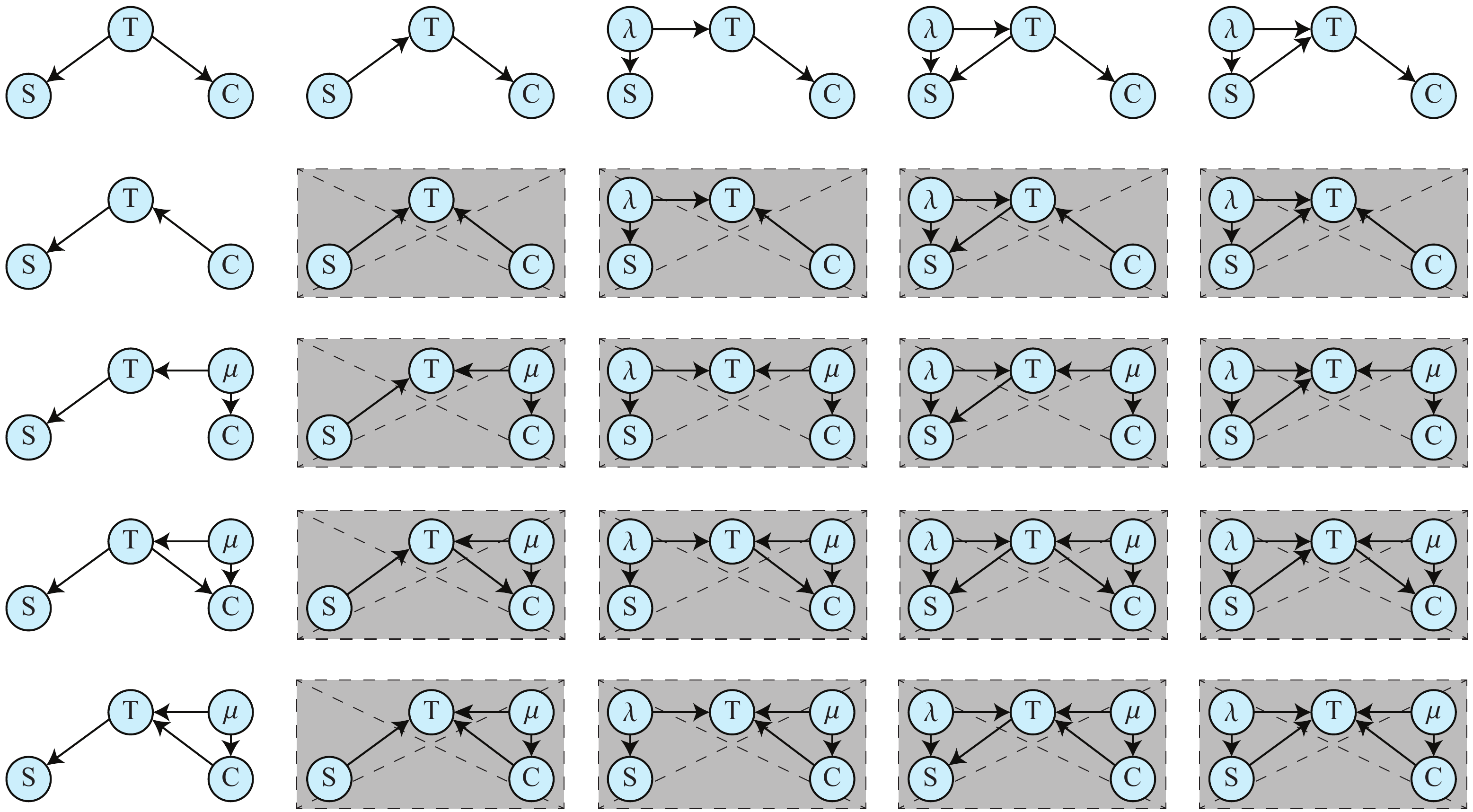}
 \caption{The causal structures returned by the IC* algorithm when the input is a distribution over observed variables $S$, $T$ and $C$ with $(S \indep C|T)$. Those that introduce a new v-structure are shaded out.}
       \label{fig:smoke-icstar-all}
\end{figure*}

For each undirected edge in this pattern, there are five possibilities in the DAG for what connection holds between the nodes, as displayed in Fig.~\ref{fig:XY-equiv}. In Fig.~\ref{fig:smoke-icstar-all} we display all twenty-five combinations of such possibilities.  We have also shaded out each of the combinations that introduces a new v-structure -- these combinations are \emph{not} candidates for the causal structure according to the IC* algorithm.  Hence, the nine causal structures that remain are the candidates returned by IC*.

How does this answer embody the principles of causal discovery?  
First, the fact that one unpacks the pattern into causal structures with only pairwise common causes is a consequence of the minimality assumption, 
as we discussed at the beginning of this section.  This is the reason that we do not find in the output of the algorithm any latent variable that is a common cause of all \emph{three} variables $S$, $T$ and $C$.

Now consider the question of why there is neither a direct causal influence between $S$ and $C$ nor a latent variable that acts as a common cause for the pair.  The  answer is simply that if either of these sorts of influences were acting, then we would \emph{not} find $(S \indep C |T)$; learning $S$ \emph{would} teach us something about $C$ even though $T$ is known. In the context of our example, this eliminates the possibility put forward by the tobacco companies of a hypothetical genetic factor that both predisposes people to smoke and to get lung cancer.

We need not consider the cases where there is also no connection between $S$ and $T$ nor the cases where there is also no connection between $T$ and $C$ because by assumption $(S \indep C|T)$ is the \emph{only} CI relation and therefore $(S \not\indep T)$ and $(T \not\indep C)$.

It follows that the twenty-five structures displayed in Fig.~\ref{fig:smoke-icstar-all} are the only possibilities that remain among all possible causal structures with pairwise common causes.  So, to explain why the output of the algorithm is justified we need only explain why one should eliminate those that introduce a new v-structure.
First note that if one conditions on a variable that is the common effect of two other variables, then we expect a dependence between those variables (for instance, in digital logic, knowing that the output of an AND gate is 0 implies that the two inputs cannot both be 1). Therefore for each causal structure that includes a v-structure on $T$, we would expect that conditioning on $T$ induces a dependence between the roots of the v-structure, and because one of these roots is always correlated with $S$ and the other with $C$, this would imply a dependence between $S$ and $C$, contradicting the fact that $(S \indep C|T)$.  Alternatively, we can infer that a causal structure including a v-structure on $T$ contradicts the relation $(S \indep C|T)$ using the d-separation criterion.

\begin{figure}[h]
		\hspace{4em}
	        \begin{subfigure}[b]{0.2\textwidth}
                	\centering
                	\includegraphics[width=\textwidth]{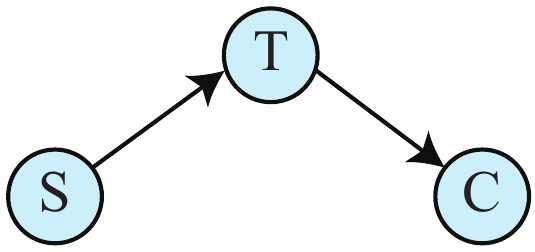}
		\end{subfigure}
		\hspace{5em}		
		\begin{subfigure}[b]{0.2\textwidth}
                	\centering
                	\includegraphics[width=\textwidth]{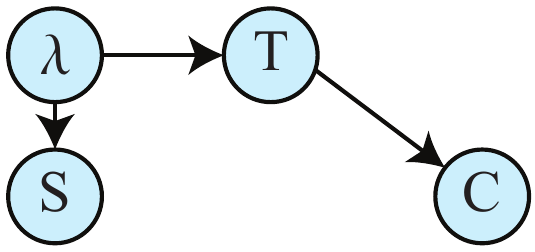}
		\end{subfigure}
		\hspace{5em}
		\begin{subfigure}[b]{0.2\textwidth}
                	\centering
                	\includegraphics[width=\textwidth]{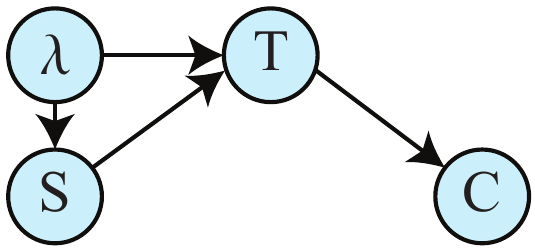}
		\end{subfigure}
 \caption{The causal structures that remain if the ordering $S<T$ is assumed.}
       \label{fig:smoke-icstar-ord}
\end{figure}

What does this imply about whether smoking causes lung cancer?  Suppose that we make use of the same additional information as we considered in Sec.~\ref{sec:causal-nohvar}, namely, that tar in the lungs is always found to occur \emph{after} smoking, never before.  We can then eliminate all causal structures with an arrow from $T$ to $S$.  What remains are the three options in Fig.~\ref{fig:smoke-icstar-ord}.  They are: (i) smoking causes tar in the lungs which causes cancer, (ii) there is a latent variable that is a common cause of smoking and having tar in the lungs, and (iii) both mechanisms are in play.  If option (ii) holds then smoking is \emph{not} a cause of cancer and, unlike the hypothesis of a genetic factor that predisposes people both to smoke and to develop lung cancer, it \emph{is} consistent with the observation that tar screens off smoking from cancer.  Of course, this hypothesis remains implausible if one cannot identify (or imagine) any factor that screens off smoking from tar in the lungs.

We previously highlighted the fact that the causal discovery algorithms of Refs.~\cite{Pearl2009} and \cite{Spirtes2001} apply the principle of minimality within the set of causal structures that are consistent with the CI relations in the observed distribution, not within the set of those that are consistent with the observed distribution itself.  This can be a problem because these two sets of causal structures can be different~\cite{Verma1993}.

\begin{figure}[h]
	        \begin{subfigure}[b]{0.2\textwidth}
                	\centering
                	\includegraphics[width=\textwidth]{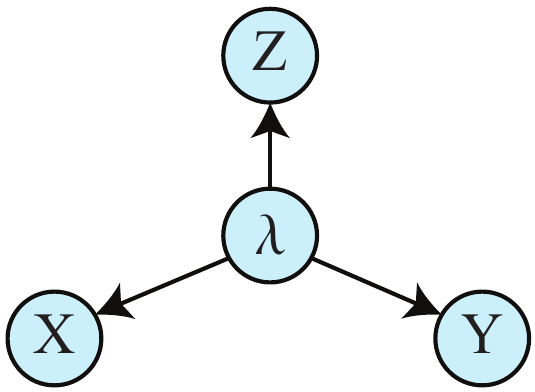}
		\end{subfigure}
		\hspace{8em}
		\begin{subfigure}[b]{0.2\textwidth}
                	\centering
                	\includegraphics[width=\textwidth]{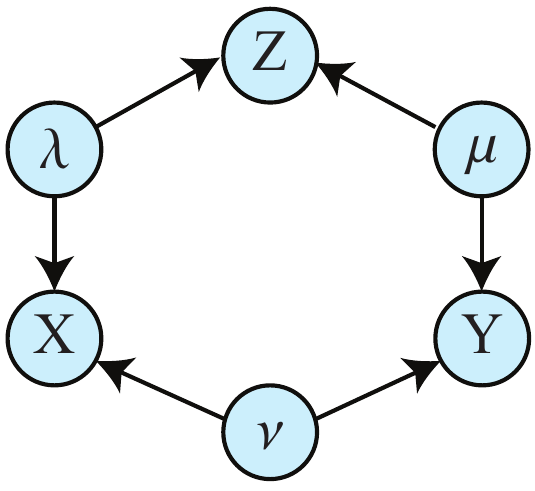}
		\end{subfigure}
 \caption{Two candidate causal structures for explaining correlations between $X$, $Y$ and $Z$ using latent variables.}
       \label{fig:XYZ-hvar}
\end{figure}

It is best to illustrate this with an example.  Consider the case of a triple of observed variables, $X$, $Y$ and $Z$.  We will compare two causal models.  The first posits a latent variable $\lambda$ which has a direct causal influence on all three observed variables.  The second posits three latent variables, $\lambda$, $\mu$ and $\nu$, each of which has a direct causal influence on a distinct pair of observed variables~\footnote{This causal scenario has also been considered in the context of a discussion of quantum correlations in Refs.~\cite{Branciard,Fritz2012}.}.  The two models are illustrated in Fig.~\ref{fig:XYZ-hvar}.

The two structures imply precisely the same set of CI relations among the observed variables, namely, the null set.
However, there are distributions over the triple of observed variables that are only consistent with the first model and not the second.  For instance, a joint distribution wherein the three observed variables $X$, $Y$ and $Z$ are close to perfectly correlated\footnote{We cannot take the case where they are perfectly correlated because we want our example to be of a distribution that is faithful to the first causal structure and perfect correlation would imply that any two variables are conditionally independent given the third.} cannot be generated from the second causal structure for any choice of causal parameters~\cite{SteudelAy,Fritz2012}.  Therefore, if this is the distribution one has observed, then the second causal structure is not a candidate for the underlying causal model.  However, the CI relations one observes for such a distribution \emph{are} consistent with the second causal structure.  
So if the input to one's causal discovery algorithm is limited to these relations, then the algorithm can return a causal structure that is inconsistent with the observed distribution. Indeed, because the first causal structure can simulate the second, the principle of minimality would naturally lead one to prefer the second, even though it is inconsistent with the observed distribution. 

We will see that this deficiency of CI-based causal discovery algorithms becomes manifest when one applies them to correlations that violate a Bell inequality.

\section{Applying causal discovery algorithms to quantum correlations}
\label{sec:qcor}

We now turn to the question of what these algorithms tell us about quantum correlations.
We consider only Bell-type experiments involving two systems, two possible settings for each measurement and two possible outcomes for each measurement.
Let $S$ and $T$ be the binary variables that specify which measurement was
performed on the left and right wings of the experiment respectively, and
let $A$ and $B$ be the binary variables that specify the outcomes of the
measurements on the left and right wings respectively.

\begin{figure}[h]
	\centering
	\includegraphics[width=0.18\textwidth]{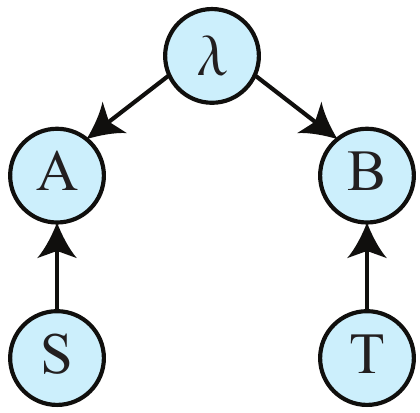}
 \caption{The causal structure corresponding to Bell's notion of local causality.
 }
       \label{fig:localcausality}
\end{figure}

Bell's theorem derives constraints
on $P(AB\,|\,ST)$ from assumptions about the causal structure~\cite{Bell1964}.  These assumptions --- which Bell justified by appeal to the space-like separation of the two wings of the experiment and the impossibility of superluminal causal influences --- are that $A$ is the joint effect of the setting variable $S$ and a common cause variable
$\lambda ,$ while $B$ is the joint effect of the setting variable $T$ and $\lambda .$ The causal structure corresponding to this assumption is presented in Fig.~\ref{fig:localcausality}.

This structure implies the following conditional independence relations,

\begin{equation*}
(A\indep BT\,|\,S\lambda) \text{ and }(B\indep AS\,|\,T\lambda).
\end{equation*}

Bell called his assumption \emph{local causality} and formalized it in terms
of these conditional independences.  These in turn imply that $
P(AB\,|\,ST\lambda )=P(A\,|\,S\lambda )P(B\,|\,T\lambda ),$ which is known as
factorizability. \ From this condition, together with the assumption that
there are no correlations between the settings and the hidden variables,%
\begin{equation*}
(S\indep T\lambda) \text{ and } (T\indep S\lambda) ,
\end{equation*}
one can infer that $P(AB\,|\,ST)$ must satisfy the Bell
inequalities~\cite{Bell1964,CHSH1969}. Bell's assumption about the causal structure also implies no superluminal signalling:
\begin{equation}\label{eq:nosignalling}
\text{No-signalling: } (A\indep T\,|\,S)\text{ and }(B\indep S\,|\,T).
\end{equation}

The fact that quantum correlations can violate Bell inequalities shows that they cannot be explained using the causal structure of Fig.~\ref{fig:localcausality}.

We will now consider the inverse problem to the one considered by
Bell. Rather than attempting to infer constraints on correlations from
assumptions about the causal structure, we will attempt to infer conclusions
about possible causal structures from the nature of the correlations that arise in quantum theory. This is the sort of problem that the causal discovery
algorithms were designed to solve.

We will contrast two examples of quantum correlations: one which violates
the Bell inequalities and the other which satisfies the Bell inequalities.

\begin{figure}[h]
	        \begin{subfigure}[b]{0.18\textwidth}
                	\centering
                	\includegraphics[width=\textwidth]{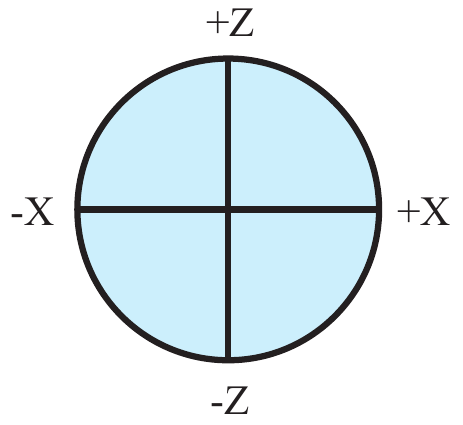}
		\caption{Left wing measurement}
		\label{fig:alice-epr}
	\end{subfigure}
	\hspace{8em}
        \begin{subfigure}[b]{0.18\textwidth}
                	\centering
                	\includegraphics[width=\textwidth]{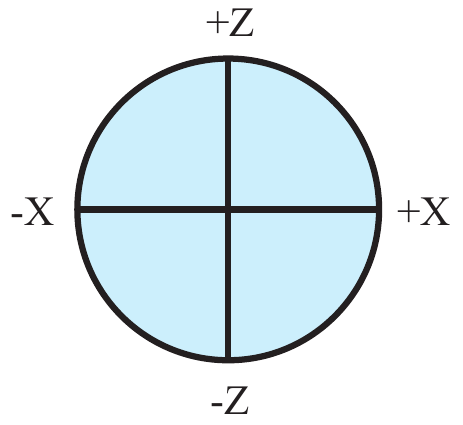}
             	\caption{Right wing measurement}
                	\label{fig:bob-epr}
        \end{subfigure}
 \caption{Measurement axes for generating EPR correlations given the quantum state $\left\vert \Psi \right\rangle$ of Eq.~(\ref{eqn:maxent})}
\label{fig:epr-meas}
\end{figure}

For the latter, we will take a version of the Einstein-Podolsky-Rosen (EPR)
experiment \cite{Einstein1935} in terms of qubits (first proposed by Bohm for spin-1/2 systems \cite{Bohmtextbook}).
The pair are prepared in the maximally entangled state
\begin{equation}
\left\vert \Psi \right\rangle =\frac{1}{\sqrt{2}}\left( \left\vert +z\right\rangle \left\vert
+z\right\rangle +\left\vert -z\right\rangle \left\vert -z\right\rangle
\right)
\label{eqn:maxent}
\end{equation}
where $\left\vert \pm z\right\rangle $ are the eigenstates of
spin along the $\hat{z}$ axis. On each wing, the two choices of measurement
are between a pair of mutually unbiased bases (the same pair for each wing).  For instance, we may measure spin along the $\hat{z}$ or $\hat{x}$ axes, as illustrated
in Fig.~\ref{fig:epr-meas}. In this case, if the same measurement is made on both wings
(both $\hat{z}$ or both $\hat{x}$), one sees perfect correlation between the
outcomes, while if \emph{different }measurements are made ($\hat{z}$ on one
and $\hat{x}$ on the other), then one sees no correlation between the
outcomes.   \ It is well known that these sorts of correlations \emph{do not
}violate any Bell inequality, which is to say that they can be explained by
a locally causal model.

\begin{figure}[h]
	        \begin{subfigure}[b]{0.18\textwidth}
                	\centering
                	\includegraphics[width=\textwidth]{alice}
		\caption{Left wing measurement}
		\label{fig:alice-chsh}
	\end{subfigure}
        \hspace{8em}
        \begin{subfigure}[b]{0.18\textwidth}
                	\centering
                	\includegraphics[width=\textwidth]{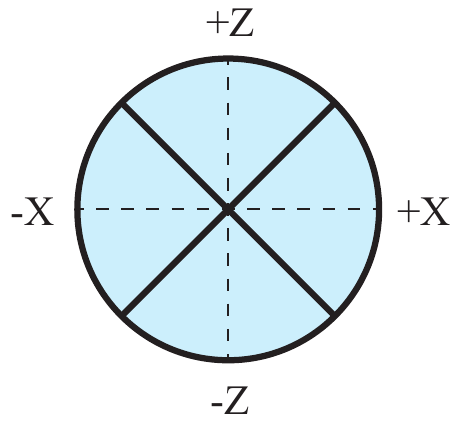}
             	\caption{Right wing measurement}
                	\label{fig:bob-chsh}
        \end{subfigure}
 \caption{Measurement axes for generating CHSH correlations given the quantum state $\left\vert \Psi \right\rangle$ of Eq.~(\ref{eqn:maxent})}
\label{fig:chsh-meas}
\end{figure}

The other sort of correlation we consider will be those exhibited in the Clauser-Horne-Shimony-Holt (CHSH) experiment. We can take the pair of spins to be prepared in the same maximally entangled state $%
\left\vert \Psi \right\rangle $ as for the EPR scenario, and the pair of
measurements on the left wing to also be of spin along the $\hat{z}$ or $%
\hat{x}$ axes. However, on the right wing, the pair of possible
measurements are of spin along the $(\hat{z}+\hat{x})/\sqrt{2}$ axis or along the $(\hat{
z}-\hat{x})/\sqrt{2}$ axis, as indicated in Fig.~\ref{fig:chsh-meas}. In this case, one finds that the probability of correlation for the cases $\left(
S,T\right) =(0,0),(1,0)$ and $(0,1)$ is equal to the probability of
anticorrelation for the cases $\left( S,T\right) =(1,1)$ and has the value $%
\frac{1}{2}+\frac{1}{2\sqrt{2}}\simeq 0.85.$

In both the EPR and CHSH scenarios, we assume that the settings $S$ and $T$ are sampled independently.

The input to the standard causal discovery algorithms is limited to
conditional independence relations, so we begin by computing the conditional
independences that hold for the EPR and CHSH experiments. Rather than
specifying an exhaustive list, we provide a generating set (the rest can be
obtained by applying the semi-graphoid axioms). They are:%
\begin{eqnarray*}
\text{EPR:} 	&&	(S\indep T),	(A\indep T\,|\,S),	(B\indep S\,|\,T),	
				(A\indep S), (B\indep T).				\\
\text{CHSH:}	&&	(S\indep T), 	(A\indep T\,|\,S),	(B\indep S\,|\,T), 
				(A\indep S), (B\indep T).			
\end{eqnarray*}%

Consider the conditions $(A\indep S)$ and $(B\indep T)$.
 These assert that the outcome on a wing is independent of the
setting on that wing. While true, this independence is not representative
of the causal structure. Indeed, it only holds because of the degeneracy
of the Schmidt coefficients in the maximally entangled state. If we instead consider the state
\[
\left\vert \Psi \right\rangle =\sqrt{p}\left\vert +z\right\rangle \left\vert
+z\right\rangle +\sqrt{1-p}\left\vert -z\right\rangle \left\vert
-z\right\rangle
\]
where $p\neq 1/2$, then $(A \not\indep S)$ and $(B \not\indep T)$.
Because it is intuitively clear that the choice of measurement \emph{does} have a causal
influence on the outcome, the independences $(A\indep S)$ and $(B\indep T)$ are pathological in the context
of the causal discovery algorithms. Given that if the EPR (CHSH)
experiment is implemented with a state that is \emph{close to} maximally entangled, it still satisfies
(violates) the Bell inequalities, we consider these states instead.  (If one
likes, $p$ may be taken to be arbitrarily close to $1/2$.)
We then get the following generating sets of independence relations,
\begin{eqnarray*}
\text{EPR: }	&&	\left( S\indep T\right) ,\left( A\indep T\,|\,S\right) ,\left( B\indep S\,|\,T\right), \label{EPRCIrelations}\\
\text{CHSH: }	&&	\left( S\indep T\right) ,\left( A\indep T\,|\,S\right) ,\left( B\indep S\,|\,T\right), \label{CHSHCIrelations}
\end{eqnarray*}%
where $\left( S\indep T\right) $ asserts the independence of the
settings, and $\left( A\indep T\,|\,S\right) $ and $\left( B\indep S\,|\,T\right) $
are the no-signalling conditions (Eq.~\eqref{eq:nosignalling}).  

The critical point is that the set of independences are the \emph{
same }for the EPR and the CHSH experiments. Since the input to the
causal discovery algorithms that we consider is limited to conditional
independence relations, it follows that whatever causal conclusions these
algorithms draw, they will draw the \emph{same} causal conclusions about the
EPR experiment as they do about the CHSH experiment. And yet, from the fact
that the EPR correlations satisfy the Bell inequalities, we know that they
\emph{can }be explained by local causes while from the fact that the CHSH
correlations violate a Bell inequality, we know that they cannot be so
explained.

So the conclusion is that CI-based causal discovery algorithms do not  do justice to Bell's theorem.
Independences simply do not provide enough information. One needs a causal
discovery algorithm that looks at the {\em strength of correlations} to reproduce Bell's conclusion.

Despite the inability of the standard causal discovery algorithms to
distinguish correlations that violate the Bell inequalities from those that
satisfy them, it is nonetheless interesting to see what happens when one
applies the algorithms to the set of independences we found for the EPR and
CHSH experiments. We will refer to these as \emph{nontrivial no-signalling
correlations} (`nontrivial' in the sense that there is some nonvanishing correlation between the outcomes
for some choices of the settings).

In applying the causal discovery algorithms, we will assume for the moment that the setting variable on one wing is a cause of the outcome variable on that wing, that is, we will assume that $S$ is a cause of $A$ and that $T$ is a cause of $B$.  This assumption will be relaxed in Sec.~\ref{sec:theorem}.
In this case, the assumption that there are no causal cycles then implies that there can be no causal influence from $A$ to $S$, nor from $B$ to $T$. Nonetheless, we are still permitting influences from the outcome on one wing to the setting on the other, although, as we will see, the causal discovery algorithms will rule against such influences.

\subsection{No latent variables}
\label{sec:qcor-nohvar}

\begin{figure}[h]
	\centering
	\includegraphics[width=0.18\textwidth]{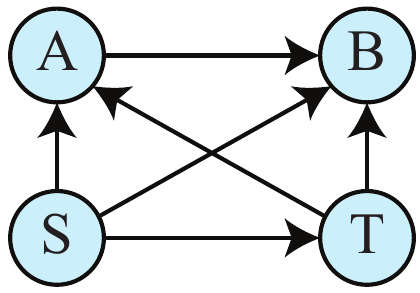}
 \caption{The most general causal structure for the causal ordering $S<T<A<B$, assuming no hidden variables.}
       \label{fig:bell-gen}
\end{figure}

It is instructive to consider the causal structure that arises for a single representative
causal ordering of the variables. We take $S<T<A<B.$ Then, the most general
causal structure is illustrated in Fig.~\ref{fig:bell-gen}. Hence the most general joint
distribution for this ordering is of the form
\begin{equation*}
P(S,T,A,B)=P(S)P(T|S)P(A|S,T)P(B|S,T,A).
\end{equation*}%

\begin{figure}[h]
        \begin{subfigure}[b]{0.18\textwidth}
                	\centering
                	\includegraphics[width=\textwidth]{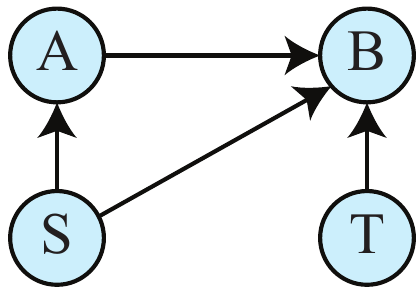}
		\subcaption{}
		\label{fig:bell-gen-AB}
	\end{subfigure}
        \hspace{8em}
        \begin{subfigure}[b]{0.18\textwidth}
                	\centering
                	\includegraphics[width=\textwidth]{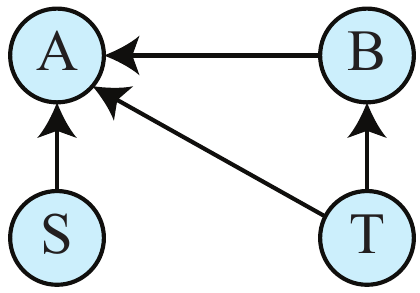}
             	\subcaption{}
                	\label{fig:bell-gen-BA}
        \end{subfigure}
        \caption{Possible causal structures for no-signalling correlations, assuming no hidden variables, for causal orderings $S<T<A<B$, $T<S<A<B$, $S<A<T<B$ (\ref{fig:bell-gen-AB}) and $S<T<B<A$, $T<S<B<A$, $T<B<S<A$ (\ref{fig:bell-gen-BA}).}
        \label{fig:bell-gen-ind}
\end{figure}

The independence $\left( S\indep T\right) $ implies that $P(T|S)=P(T),$ and
the independence $\left( A\indep T\,|\,S\right) $ implies that $P(A|S,T)=P(A|S)$.
The independence$\ \left( B\indep S|T\right) $ has no nontrivial implications
for this causal ordering, hence the term $P(B|S,T,A)$ cannot be simplified. From these CI relation it
follows that the joint distribution can be written as
\begin{equation*}
P(S,T,A,B)=P(S)P(T)P(A|S)P(B|S,T,A),
\end{equation*}%
which corresponds to the causal structure in Fig.~\ref{fig:bell-gen-AB}. If we change the
ordering of variables so that $B$ precedes $A,$ then by a similar argument,
we obtain the causal structure in Fig.~\ref{fig:bell-gen-BA}.  For every other possible causal ordering consistent with our assumption that $S<A$ and $T<B$, we also obtain one of the causal structures of Fig.~\ref{fig:bell-gen-ind}.

Consider the causal structure in Fig.~\ref{fig:bell-gen-AB}, Although it faithfully
captures $\left( S\indep T\right) $ and $\left( A\indep T|S\right) ,$ it does
not faithfully capture $\left( B\indep S|T\right) .$ The only way to
explain the independence $\left( B\indep S|T\right) $ within this causal
model is by fine-tuning of the causal parameters in the model, for instance,
if the parameters defining $P(B|S,T,A)$ are not independent of those
defining $P(A|S)$. A similar problem arises for the causal structure in
Fig.~\ref{fig:bell-gen-BA}. It follows that in the case of no latent variables, no
causal structure can satisfy faithfulness for the conditional independences of nontrivial
no-signalling correlations.

Note that if, instead of applying the Wermuth-Lauritzen algorithm to the nontrivial no-signalling correlations, one applies the IC algorithm~\cite{Pearl2009}, equivalently the SGS algorithm~\cite{Spirtes2001} (which also assume no latent variables), one finds that it returns a graph that is not a valid pattern, signalling a failure of the algorithm.  This is what one would expect given that the algorithm only promises to return a valid causal structure if there exists one that satisfies faithfulness, and in this case, there is not.

There is an interesting lesson here for the foundations of quantum theory.
Long before Bell's work, Einstein had pointed
out that if one did \emph{not} assume hidden variables, then one could only
explain the EPR correlations by positing superluminal causes. This
argument was made in his comments at the 1927 Solvay conference~\cite{Bacci2009} (See Refs.~\cite{Norsen2010} and ~\cite{Harrigan2010} for more concerning Einstein's
arguments on completeness and locality.) One can easily cast Einstein's
argument into the mold of causal discovery algorithms as follows. If we allow the quantum state $\psi$, considered as a classical variable, as the only common cause, then the assumption of no superluminal
causal influences implies that $P(A,B|S,T,\psi )=P(A|S,\psi )P(B|T,\psi )$, and given that $
\psi $ is fixed in the experiment (it is a variable which only takes one
possible value), this implies that $A$ and $B$ should be uncorrelated, in
contradiction with the EPR correlations.

But Einstein failed to explicitly note another mysterious feature of the EPR correlations, which our analysis highlights:
\emph{even if} one is willing to countenance superluminal causal influences in an
attempt to explain the EPR correlations without recourse to hidden variables,
ensuring that these superluminal causes cannot be used to send
superluminal signals implies that there must be fine-tuning in the
underlying causal model.

\subsection{Latent variables allowed}
\label{sec:qcor-hvar}

\begin{figure}[h]
	\centering
	\includegraphics[width=0.18\textwidth]{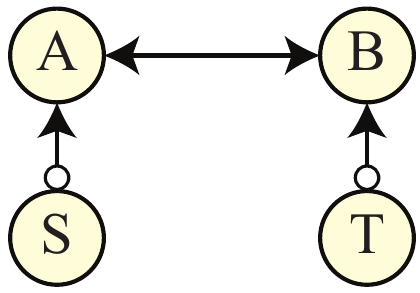}
 \caption{The output pattern of the IC* algorithm when applied to nontrivial no-signalling correlations.}
       \label{fig:bell-ics}
\end{figure}

If one simply inputs the independences of nontrivial no-signalling
correlations into the IC* algorithm of Ref.~\cite{Pearl2009}, which allows latent variables, one obtains the pattern illustrated in Fig.~\ref{fig:bell-ics} as output.

Recall that the arrows with an empty circle at their tail imply
that one can have either a direct causal link or a common cause.
If one believes that the settings at each wing are freely chosen, then one
is inclined to think that either the setting variables $S$ and $T$ should
be direct causes of $A$ and $B$ respectively, or that if they are not, then it is
the common cause for $A$ and $S$ and the common cause for $B$ and $T$ that is freely chosen.
In this case, we could lump the common causes into the definition of the setting variables without loss of generality.

Besides this caveat about the causal relation between $S$ and $A$ and between
$T$ and $B,$ the causal structures with pairwise common causes that are consistent with the pattern that the IC* algorithm has returned are
precisely those that capture Bell's notion of local causality, illustrated in Fig.~\ref{fig:localcausality}.  Moreover, the principle of minimality, applied to the causal models consistent with the CI relations, would lead us to favour the causal model of Fig.~\ref{fig:localcausality}.    But because such a causal model satisfies the Bell inequalities, while the CHSH correlations do not, we know that it cannot provide a causal explanation of the CHSH correlations. 

This is how the deficiency of the IC* algorithm manifests itself when applied to quantum correlations.    The problem is that a causal structure with latent variables that reproduces the CI relations of a given distribution might not be capable of reproducing the distribution itself.  In particular, the causal structure of Fig.~\ref{fig:localcausality} reproduces the CI relations of the distribution $P(A,B,S,T)$ defined by the CHSH experiment, namely $\left( S\indep T\right) ,\left( A\indep T\,|\,S\right)$ and $\left( B\indep S\,|\,T\right)$, but it cannot reproduce the distribution itself.  
As our brief discussion in Sec.~\ref{sec:causal-hvar} highlighted, if one applies the principle of minimality among the causal models {\em that are consistent with the CI relations}, rather than among the causal models {\em that are consistent with the entire observed distribution}, one can mistakenly come to favour a causal model that cannot reproduce the observed distribution.

Of course, we already pointed out in Sec.~\ref{sec:qcor}, that the input of the IC*
algorithm cannot distinguish Bell-inequality-violating from
Bell-inequality-satisfying correlations.  
So we reiterate our conclusion from Sec.~\ref{sec:qcor}, that causal discovery
algorithms which look only at conditional independences are inadequate to the task of
establishing whether or not correlations can be explained by a locally
causal model. We require better algorithms that also take into account the strengths of the correlations.

\subsection{Some proposed causal explanations of quantum correlations}

We now apply the ideas behind causal discovery algorithms to a few of the existing proposals for providing a causal explanation of Bell-inequality-violating correlations.  We consider three: superluminal causation, superdeterminism, and retrocausation.

We start by considering the most general kind of causal explanation here, where one allows hidden variables.  Causal structures \emph{without} hidden variables are a special case of these. Nonetheless, we consider the case of no hidden variables explicitly to ensure that there is no confusion.

\subsubsection{Superluminal causation}

\begin{figure}[h]
        \begin{subfigure}[b]{0.18\textwidth}
                	\centering
        	\includegraphics[width=\textwidth]{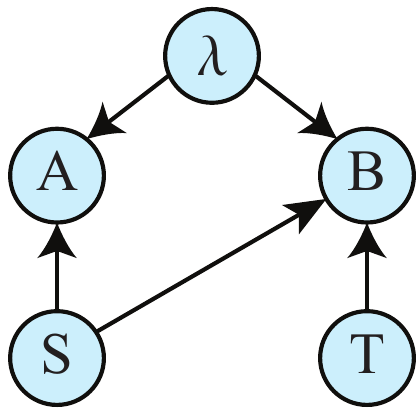}
		\subcaption{}
		\label{fig:bell-StoB}
	\end{subfigure}
	\hspace{5em}
        \begin{subfigure}[b]{0.18\textwidth}
                	\centering
        	\includegraphics[width=\textwidth]{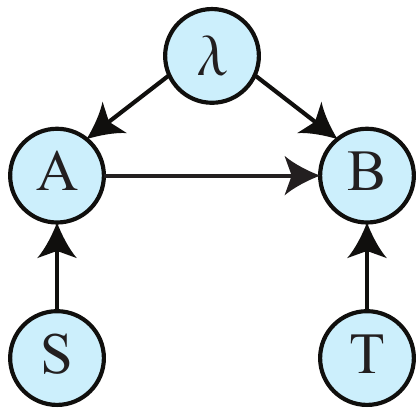}
             	\subcaption{}
                	\label{fig:bell-AtoB}
        \end{subfigure}
        \hspace{5em}
                \begin{subfigure}[b]{0.18\textwidth}
                	\centering
        	\includegraphics[width=\textwidth]{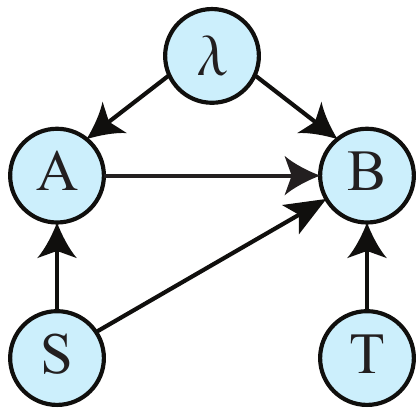}
             	\subcaption{}
                	\label{fig:bell-SAtoB}
        \end{subfigure}
 \caption{Examples of causal structures that posit superluminal causal influences to explain Bell correlations.}
\label{fig:bell-SandAtoB}
\end{figure}

One option for explaining Bell correlations causally is to assume that there are some superluminal causes, for instance, a causal influence from the outcome on one wing to the outcome on
the other, or from the setting on one wing to the outcome on the other, or both. 
  The possibilities are illustrated in Fig.~\ref{fig:bell-SandAtoB}.

 These sorts of causal explanations of Bell-inequality violations, however, are unsatisfactory in light of the principles embodied in causal discovery algorithms.  Given the superluminal \emph{causal influences} from one wing to the other, the only way to explain the lack of superluminal \emph{signals}, that is the CI relations of Eq.~\eqref{eq:nosignalling}, is through a fine-tuning of the causal parameters.\color{black}

For instance, in Fig.~\ref{fig:bell-SAtoB}, the correlations set up between $S$ and $B$ along the direct causal path could cancel with those set up by the causal path through $A$. (The path through $\lambda$ cannot set up correlations between $S$ and $B$ because there is a collider on $A$ in this path and we are not conditioning on $A$.) Such a cancelation requires fine-tuning of the parameters of the model.

To salvage no-signalling for the causal structure of Fig.~\ref{fig:bell-StoB}, we need a different sort of fine-tuning (a similar sort of fine-tuning mechanism can also be used for the causal structure of Fig.~\ref{fig:bell-AtoB}).  For instance, it could be that $\lambda=(\lambda_1,\lambda_2)$ where $\lambda_1$ is a binary variable that is uniformly distributed and that $B$ is a function of $S\oplus \lambda_1$, $T$ and $\lambda_2$. In this case, we can ensure that $(B \indep S|T)$ by virtue of the special distribution on $\lambda_1$, which is a kind of fine-tuning.

Note that this is precisely the sort of causal structure that is assumed in the Toner and Bacon model~\cite{TonerBacon2003}, where Bell-inequality violations are simulated by classical communication\footnote{This model works even when the measurement setting for each qubit is chosen arbitrarily, rather than being limited to the two settings of the CHSH experiment.}. This model also involves fine-tuning insofar as signalling is prohibited only for a special distribution over the shared random variables posited by the model.

The deBroglie-Bohm interpretation is a prominent example of a model that seeks to provide a causal explanation of Bell correlations using superluminal causal influences. Consider the
deBroglie-Bohm interpretation of a relativistic theory such as the model of QED provided by Struyve
and Westman~\cite{Struyve2007}, or else of a nonrelativistic theory wherein the interaction
Hamiltonians are such that there is a maximum speed at which signals can
propagate.  In both cases, it is presumed that there is a preferred rest frame that is hidden at the operational level.
In a Bell experiment, if the measurement on the left wing occurs prior to the measurement on the right wing relative to the preferred rest frame, then there is a superluminal causal influence from the setting on the left wing to the outcome on the right wing, mediated by the
quantum state, which is considered to be a part of the ontology of the
theory~\cite{Bohm1993}.  (Note that no causal influence from the outcome of the first experiment to the
outcome of the second is required because the outcomes are deterministic functions of the Bohmian configuration and the wavefunction.) It follows from our analysis that the parameters in the causal model posited by the deBroglie-Bohm interpretation must be fine-tuned in order to explain the lack of superluminal signalling.

Valentini's version of the deBroglie-Bohm interpretation makes this fact
particularly clear. In Refs.~\cite{Valentini1991a,Valentini1991b} he has noted that the
wavefunction plays a dual role in the deBroglie-Bohm interpretation. On the
one hand, it is part of the ontology, a pilot wave that dictates the
dynamics of the system's configuration (the positions of the particles in
the nonrelativistic theory). On the other hand, the wavefunction has a statistical character, specifying the  distribution over the system's configurations. In order to
eliminate this dual role, Valentini suggests that the wavefunction is
only a pilot wave and that \emph{any} distribution over the configurations
should be allowed as the initial condition. It is argued that one
can still recover the standard distribution of configurations on a coarse-grained scale
as a result of  dynamical evolution~\cite{Valentini2008}. Within this approach, the no-signalling
constraint is a feature of a special equilibrium distribution.
The tension between Bell inequality violations and no-signalling is resolved
by abandoning the latter as a fundamental feature of the world and asserting
that it only holds as a \emph{contingent} feature. The fine-tuning is
explained as the consequence of equilibration. (It has also been noted in the
causal model literature that equilibration phenomena might account for fine-tuning of causal parameters~\cite{dash2005}.) 

Conversely, the version of the deBroglie-Bohm interpretation espoused by D\"{u}rr, Goldstein and Zhangi~\cite{Durr1992}  -- which takes no-signalling to be a non-contingent feature of the theory -- does not seek to provide a dynamical explanation of the fine-tuning. Consequently, it seems fair to say that the fine-tuning required by the deBroglie-Bohm interpretation is less objectionable in Valentini's version of the theory. On the other hand, the cost of  justifying the fine-tuning by a dynamical process of equilibration is that, because true equilibrium is an idealization that is never achieved in finite time, one would expect systems to have small deviations from equilibrium and such deviations could in principle be exploited to send signals superluminally.  Valentini endorses this consequence of his version of the deBroglie-Bohm interpretation~\cite{Valentini2002} and indeed has made proposals for where the strongest deviations from equilibrium might arise~\cite{Valentini2004}.  Therefore, anyone who thinks that the absence of superluminal signals is a necessary, rather than a contingent, feature of quantum theory, will not be enthusiastic about Valentini's approach.

Another recent article that considers the question of whether Bell-inequality violations can be explained by superluminal causal influences is Bancal {\it et al.}~\cite{Bancal2012}.  They consider a physical model wherein causal influences can propagate at a speed $v$ that is faster than light, $v >c$, but still finite.  They imagine that this model deviates from quantum theory in some of its operational predictions.   In particular, if the two wings of a Bell experiment are space-like separated relative to the $v$ lightcone structure, then it is presumed that the Bell inequalities are not violated (contrary to the quantum predictions), whereas if they are time-like separated relative to $v$ (but still space-like separated relative to $c$), the Bell inequalities are violated, as they would be in quantum theory.  They then show that the the superluminal causes in their model can be leveraged to achieve superluminal signalling.  In this sense, their model is analogous to Valentini's model, and implies that if one is unwilling to endorse a theory allowing such signals, one should not posit finite superluminal causal influences.  Note, however, that, unlike Valentini's model, the Bancal {\it et al.} model still requires fine-tuning in those cases wherein the superluminal causal influences cannot be used to send superluminal signals, such as the original bipartite scenario.  The fine-tuning criticism of explanations positing superluminal causes applies whether those the superluminal causes propagate at a finite speed or not.

\subsubsection{Superdeterminism}

Another option for a causal explanation of quantum correlations is to posit that the settings are not free but are causally influenced by other variables.

\begin{figure}[h]
        \begin{subfigure}[b]{0.18\textwidth}
                	\centering
        		\includegraphics[width=\textwidth]{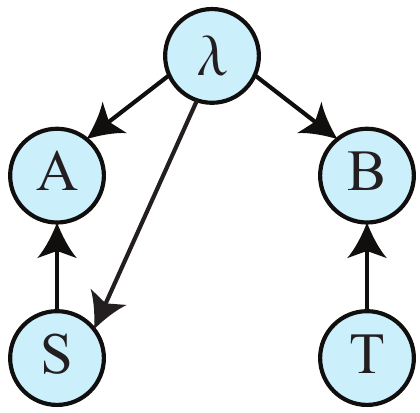}
		\subcaption{}
		\label{fig:superdeta}
	\end{subfigure}
	\hspace{5em}
        \begin{subfigure}[b]{0.18\textwidth}
                	\centering
        	\includegraphics[width=\textwidth]{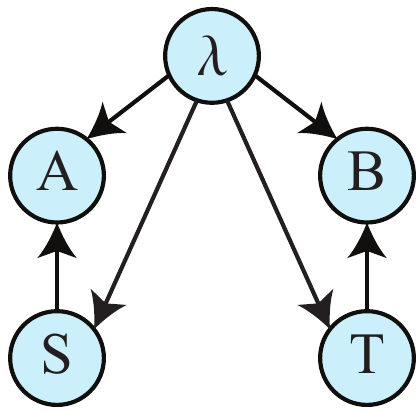}
             	\subcaption{}
                	\label{fig:superdetb}
        \end{subfigure}
        \hspace{5em}
        \begin{subfigure}[b]{0.18\textwidth}
                	\centering
        	\includegraphics[width=\textwidth]{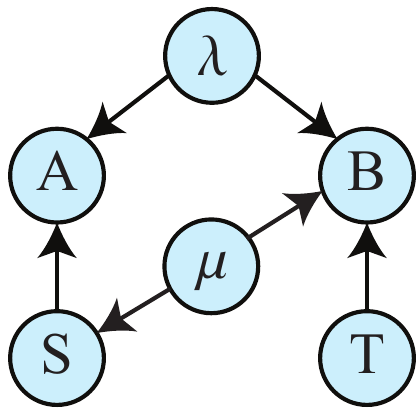}
             	\subcaption{}
                	\label{fig:superdetc}
        \end{subfigure}
 \caption{Some causal structures that exploit the superdeterminism loophole to explain Bell correlations.}
\label{fig:superdet}
\end{figure}

For instance, the hidden variable $\lambda$ (which correlates the outcomes) might causally influence one or both of the setting variables, as illustrated in Figs.~\ref{fig:superdeta} and~\ref{fig:superdetb}. Alternatively, one can posit the existence of a second hidden variable $\mu$ that is a common cause for the setting on one wing and the outcome on the other wing, as illustrated in Fig.~\ref{fig:superdetc}.  More complicated possibilities would have $\mu$ as a common cause of a subset of three of the settings and outcomes.
Note that the possibility of a latent variable that is a common cause of $\lambda$ and one or both settings has not been excluded; it is incorporated into the first case.  This is because any such variable could just be absorbed into the definition of $\lambda$ without loss of generality.
The scenario in Fig.~\ref{fig:superdetc} could also be considered a special case of the one in Fig.~\ref{fig:superdeta}, if we include $\mu$ into the definition of $\lambda$.  Nonetheless, it is useful to separate out this second case because it posits that the common cause of $A$ and $B$ is not correlated with the common cause of $S$ and $B$.

All of the causal influences posited in such models can be taken to be subluminal.  However, such explanations of the Bell correlations are clearly in conflict with the notion that the settings can be freely chosen by the experimenter. To assert one of these causal structures as a way to resolve the mystery of Bell's theorem is an instance of what is commonly known as the ``superdeterminism'' loophole.
But, just as with positing superluminal causal influences, these causal structures are not faithful to the observed correlations because one or more of the observed CI relations -- $S \indep T$ (independence of settings),  $\left( A\indep T|S\right)$ (no-signalling from left to right) and $\left( B\indep S|T\right)$ (no signalling from right to left) -- can only be satisfied by fine-tuning of the parameters in the causal model.  This is a novel sort of objection against the notion of a superdeterministic explanation of Bell-inequality-violations, independent of an appeal to free will.

It is worth devoting a few words to the sort of fine-tuning that is required.  First note that in the context of abandoning the assumption of free will, the no-signalling constraint must be reinterpreted as an observed statistical independence, rather than a statement about the consequences of an intervention on a setting variable.  Of course, this statistical independence is still observed and therefore must still be reproduced by the causal model. In the causal structure of Fig.~\ref{fig:superdeta}, if we define $\lambda^*$ to be that part of $\lambda$ that is correlated nontrivially with $S$, then we require that $\lambda^* \indep B$ despite the arrow from $\lambda$ to $B$.  We can still do justice to the Bell correlations by having $\lambda^*$ correlated with only
the \emph{parity} of $A$ and $B$, while remaining uncorrelated with $B$.  This is an instance of fine-tuning.

Similar fine-tuning tricks can be used to ensure that $(B \indep S|T)$ in the causal structures of Figs.~\ref{fig:superdetb} and ~\ref{fig:superdetc}.

\subsubsection{Retrocausation}

``Retrocausation'' refers to the possibility of causal influences that act in a direction contrary to the standard arrow of time. It has been proposed as a means of resolving the mystery of Bell-inequality violations \cite{deBeauregard1953,Cramer1986,Sutherland2006,Price2008,Wharton2010} by purportedly saving the relativistic structure of the theory: rather than having causal influences propagating outside the light cone, they propagate \emph{within} the light cone although possibly within the \emph{backward} light cone.

It is useful to distinguish two approaches to retrocausal explanations of Bell correlations: those that add cycles to the causal structure and those that do not.  Given that the former take us outside the framework of directed \emph{acyclic} graphs, we will confine our attention to acyclic retrocausation.

\begin{figure}[h]
        \begin{subfigure}[b]{0.18\textwidth}
                	\centering
        		\includegraphics[width=\textwidth]{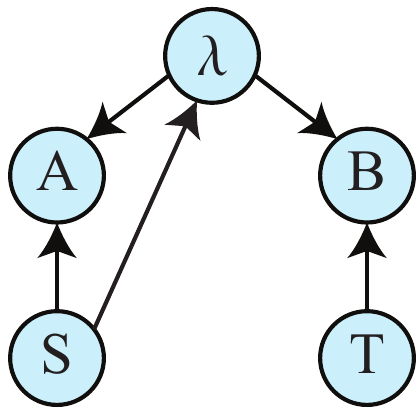}
		\subcaption{}
		\label{fig:retroa}
	\end{subfigure}
	\hspace{5em}
        \begin{subfigure}[b]{0.18\textwidth}
                	\centering
        	\includegraphics[width=\textwidth]{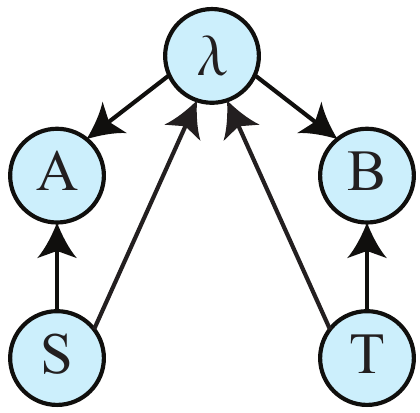}
             	\subcaption{}
                	\label{fig:retrob}
        \end{subfigure}
        \hspace{5em}
        \begin{subfigure}[b]{0.18\textwidth}
                	\centering
        		\includegraphics[width=\textwidth]{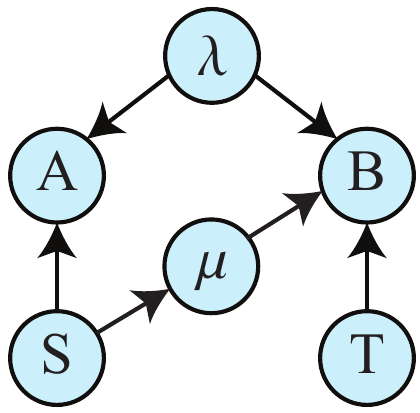}
             	\subcaption{}
                	\label{fig:retroc}
        \end{subfigure}
 \caption{Causal structures that exploit the retrocausation loophole to explain Bell correlations.}
 \label{fig:retro}
\end{figure}

Price has described the idea of a retrocausal model of Bell inequality violations in Ref.~\cite{price1997time}.  It is not completely clear whether he has in mind a model that posits cycles or not.  However, he does argue that one way to generate a retrocausal model is to start with a superdeterministic model and to simply reverse the causal arrows that lead into the settings.  For the examples of superdeterminism we have considered, such reversals lead to acyclic retrocausal models.
For instance, if one starts with the superdeterministic causal structure of Fig.~\ref{fig:superdeta} and reverses the $\lambda \to S$ arrow, one obtains the causal structure of Fig.~\ref{fig:retroa}, where setting $S$ is a cause of the hidden variable $\lambda$.  If one assumes that $S$ is chosen freely at a time to the future of when $\lambda$ is set, then this model is clearly retrocausal.

Alternatively, consider taking the superdeterministic model of Fig.~\ref{fig:superdetc} and reversing the $\mu \to S$ arrow, to obtain the causal structure of Fig.~\ref{fig:retroc}.  If $\mu$ were presumed to be space-like separated from both $S$ and $B$, it would simply mediate a superluminal causal influence from $S$ to $B$.  However, if one posits that $\mu$ is in the common future of $S$ and $B$, then we can imagine that there is a causal influence from $S$ to $\mu$ that is subluminal, and one from $\mu$ to $B$ that is retrocausal.  Alternatively, if one posits that $\mu$ is in the common past of $S$ and $B$, then the causal influence from $S$ to $\mu$ must be assumed to be retrocausal.

Note that if one views spatio-temporal relations as supervening upon causal relations, rather than vice-versa, then there is no freedom to specify the spatio-temporal location of $\mu$ and the distinction drawn above is not meaningful. Even if one takes spatio-temporal notions to be primary, the fact that the location of $\mu$ seems to be mere window-dressing in the context of a causal explanation of Bell-inequality violations undermines the distinction between retrocausation and superluminal causation.

Fine-tuning is just as necessary within the retrocausal explanations as it was in the ones that posited superluminal influences or superdeterminism.  Without it, one would obtain a correlation between $S$ and $B$, in contradiction with their observed statistical independence. Indeed, if these causal structures could be supplemented with arbitrary causal parameters, then one could use the causal chain of influence that extends from $S$ to $B$ to send a signal.


\subsubsection{Causal explanations without hidden variables}

Note that causal structures \emph{without} hidden variables can always be interpreted as causal structures with trivial hidden variables.  
If $\lambda$ is a variable acting as a common cause of other variables but takes only a \emph{single} value, then it cannot generate a statistical correlation among its causal children.  Equivalently, a variable acts as a nontrivial common cause only if the distribution over its values has a nontrivial spread.

Recall Fig.~\ref{fig:bell-SandAtoB}, which considers both a common cause relation holding between $A$ and $B$ together with a superluminal causal influence from the settings and/or outcomes on one wing to the outcome on the other wing.  If the hidden variable $\lambda$ is trivial, then we can drop it from the causal structure, to obtain the following DAGs, wherein $A$ and $B$ are related only by a superluminal causal influence from the settings and/or outcomes on one wing to the outcome on the other wing. 

\begin{figure}[h]
        \begin{subfigure}[b]{0.18\textwidth}
                	\centering
        	\includegraphics[width=\textwidth]{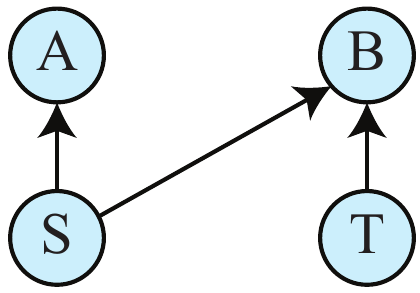}
		\subcaption{}
		\label{nohva}
	\end{subfigure}
	\hspace{5em}
        \begin{subfigure}[b]{0.18\textwidth}
                	\centering
        	\includegraphics[width=\textwidth]{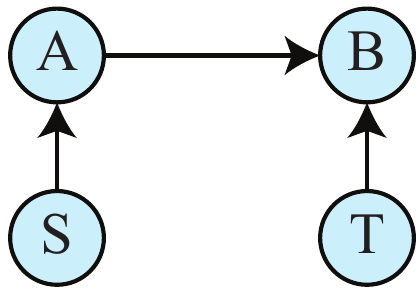}
             	\subcaption{}
	\label{nohvb}
        \end{subfigure}
        \hspace{5em}
                \begin{subfigure}[b]{0.18\textwidth}
                	\centering
                	\includegraphics[width=\textwidth]{bell-gen-AB}
		\subcaption{}
			\label{nohvc}
	\end{subfigure}
 \caption{Examples of causal structures that posit superluminal causal influences but no hidden variables to explain Bell correlations.}
\label{fig:bellnohv-SandAtoB}
\end{figure}

The first two of these causal diagrams, \ref{nohva} and \ref{nohvb}, are not viable as causal explanations of the observed correlations because they each imply a CI relation that is not observed:  \ref{nohva} implies $\left( A \indep B|S\right)$, while \ref{nohvb} implies $\left( S \indep B|A \right)$.  So only \ref{nohvc} is a candidate for a causal explanation.  However, it is obvious that this causal structure requires fine-tuning to explain the no-signalling independence $B \indep S$ just as much as the causal structures of Fig.~\ref{fig:bell-SandAtoB} do.

The fact that trivial variables as common causes are useless for explaining correlations
serves to clarify how Bell's theorem provides a challenge to the possibility of causal explanations of quantum correlations  
{\em even if} one espouses the interpretation of quantum theory wherein the pure quantum state $\psi$ is a complete description of reality (this is called a $\psi$-complete ontological model in the terminology of Ref.~\cite{Harrigan2010}).  Believing that Bell's theorem only holds if one assumes hidden variables is a common error.  We therefore pause to consider what is wrong with an intuition which may well acount for the error, namely, that in a Bell experiment the common cause of the measurement outcomes is the quantum state of the pair of particles.  

Within the framework of causal models, only a {\em variable} can act as a common cause. If one takes the quantum state of the pair of particles  to be a common cause within this framework, then it is a trivial variable: it is fixed in the experiment to some particular state $\psi$, which is known to the experimentalist.  Consequently, the quantum correlations are of the form 
\[
P(A,B|S,T)=P(A|S,\psi)P(B|T,\psi),
\]
which is a product distribution.  Alternatively, if one considers a variable $\Psi$ that varies over all possible quantum states for the pair of particles, then in the experiment the distribution over this variable is a Dirac-delta distribution centered on $\psi$, $P(\Psi)=\delta(\Psi-\psi)$, and it again follows that 
\[
P(A,B|S,T)=\int {\rm d}\Psi P(A|S,\Psi)P(B|T,\Psi) P(\Psi)= P(A|S,\psi)P(B|T,\psi).
\]

  Therefore, any causal model wherein the quantum state is considered to be a common cause does not yield {\em any} correlation between the measurement outcomes, let alone correlations that violate a Bell inequality.  
Indeed, if one does not allow hidden variables, then the only way to explain correlations between $A$ and $B$, even those that satisfy the Bell inequalities, is through a causal influence from the settings and/or outcomes on one wing to the outcome on the other wing, as in Fig.~\ref{fig:bellnohv-SandAtoB}.
But to make this influence consistent with the no-signalling independence relations, one requires fine-tuning. 
Because the possiblity of an explanation in terms of a {\rm hidden} common cause, distinct from the quantum state,
 has been ruled out from the outset by the assumption of no hidden variables, Bell's argument is not required in this case.
 This is precisely what we concluded in our discussion of EPR correlations in Sec.~\ref{sec:qcor-nohvar}.

\section{Proof of the necessity of fine-tuning in causal explanations of Bell inequality violations}
\label{sec:theorem}

Up until this point, we have followed closely the methodology of existing CI-based causal discovery algorithms.  
Such algorithms aim to contend with scenarios wherein the number of observed variables may be large.
 Our interest here, however, is Bell-type experiments, where there are only four observed variables.  
 It follows that we can ignore the details of existing algorithms and instead use a brute-force search to determine whether there is any causal explanation of the correlations observed in such experiments that can do justice to the core principles of causal discovery algorithms. 
 We find that no such causal explanation is possible.  This can be understood as a novel characterization of Bell's theorem. 

We begin by clarifying the assumptions of our  no-go theorem.

\begin{quote}
QCORR:  The assumption that the following predictions of quantum theory are correct.  In a Bell-type experiment where the settings and outcomes are binary variables, with $S$ ($T$) and $A$ ($B$) denoting, respectively, the setting and outcome on the left (right) wing, it is possible to find a probability distribution $P(A,B,S,T)$ such that: (i) the conditional independence relations $\left( S\indep T\right)$, asserting the marginal independence of the
settings, as well as $\left( A\indep T\,|\,S\right) $ and $\left( B\indep S\,|\,T\right) $, asserting the conditional independence of each local outcome from the distant setting given the local setting (which is the standard way of formalizing the assumption of no superluminal signals) are satisfied, and (ii) a Bell inequality is violated. 
\end{quote}

\begin{quote}
CAUSAL:  The assumption that a probability distribution over observed variables 
can be explained causally using the standard framework of causal models, described in Sec.~\ref{sec:causalmod}.  This framework presumes Reichenbach's principle, which asserts that any correlation between a pair of variables must be explained either by a causal influence from one variable to the other, or a common cause acting on both, or a combination of the two mechanisms.  
\end{quote}

\begin{quote}
NOFINE: The assumption that the conditional independence relations that hold in a probability distribution over observed variables 
are a consequence of the causal structure alone rather than a consequence of a particular choice of values for the causal-statistical parameters.  This is the principle of {\em faithfulness} or {\em no fine-tuning} described in Sec.~\ref{sec:causalalg}.
\end{quote}

Our no-go theorem establishes that under the assumption of the framework of causal models,
 every causal model that can reproduce no-signalling Bell-inequality-violating correlations must be fine-tuned.  Formally, it can be characterized as the inconsistency of our three assumptions above:

\begin{theorem}
CAUSAL $+$ NOFINE $+$ QCORR $\Rightarrow$ contradiction
\end{theorem}

\begin{proof}
We must consider all candidates for a causal
model underlying the correlations among the observed variables $A$, $B$, $S$ and
$T.$

Recall that there is no point adding latent variables that are transits
between observed variables or the common effect of observed variables, because
these yield the same possibilities for correlations among the observed variables as causal models that
exclude such variables. 
So we need only
consider adding latent variables that act as common causes of observed variables.

From the CI relations, $\left( S\indep T\right)$, $\left( A\indep T\,|\,S\right) $ and $\left( B\indep S\,|\,T\right)$, one can deduce the CI relations $\left(S\indep B\right)$ and $\left(T\indep A\right)$ using the semi-graphoid axioms. For instance, from $\left( S\indep T\right)$ and $\left( B\indep S\,|\,T\right)$, the contraction axiom allows one to infer $\left(S \indep BT\right)$, and then from the decomposition axiom, one obtains $\left( S\indep B\right)$.  

Consider which {\em pairs} of the observed variables could possibly
admit of a hidden common cause.  We can immediately exclude hidden common causes for the pairs $\left\{  S,T\right\}  $,
$\left\{  S,B\right\}  $ and $\left\{  T,A\right\}  $ because if there were such hidden common causes then the only way to obtain the observed
independences $\left(S\indep T\right),$ $\left(S\indep B\right)$ and $\left(T\indep A\right)$ would be by fine-tuning.  This in turn implies that there cannot be a hidden common cause for all four of the observed variables, because it would imply a hidden common cause for these pairs.  Furthermore, because any triple of the observed variables necessarily includes at least one of these pairs, there cannot be a hidden common cause for any triple either.
It follows therefore that the only sets of observed variables for which we need to consider the possibility of a hidden common cause are the pairs $\left\{A,B\right\},$ $\left\{  S,A\right\}  $ and $\left\{  T,B\right\}$.

Now consider which pairs of the observed variables could possibly be connected by a direct causal influence.  We can exclude such
influences for the pairs $\left\{S,T\right\}$,
$\left\{S,B\right\}$ and $\left\{T,A\right\}$, because if any such influence were present, we could
only ensure $\left(S\indep T\right),$ $\left(S\indep B\right)$ and $\left(T\indep A\right)$ by fine-tuning. \ The only
pairs for which there can be a direct causal influence, therefore, are $\left\{A,B\right\},$ $\left\{  S,A\right\}  $ and $\left\{  T,B\right\}.$

Finally, consider which pairs of the observed variables could possibly have no causal connection between the elements of the pair. From the fact that the pairs $\left\{S,T\right\}$, $\left\{S,B\right\}$ and $\left\{T,A\right\}$ admit neither a hidden common cause nor a direct causal influence, they are necessarily in the set of pairs for which there is no causal connection.   Conversely, we can exclude the possibility of no causal connection for the pairs $\left\{A,B\right\},$ $\left\{  S,A\right\}  $ and $\left\{  T,B\right\}$, because there would then be no way of explaining the observed correlations between $S$ and $A$ or between $T$ and $B$ or between $A$ and $B$ given $S$ and $T.$

So, nontrivial causal connections can arise only for the pairs $\left\{A,B\right\},$ $\left\{  S,A\right\}  $ and $\left\{  T,B\right\}$, and in each case there can be a direct causal influence in either direction, a hidden common cause, or a hidden common cause acting jointly with a direct causal influence in one direction; these are just the five possibilities outlined in Fig.~\ref{fig:XY-equiv}.

We now exclude all the possibilities wherein there is a direct causal influence from $A$ to $B$.  We do this by noting that whichever of the five causal connections hold between $S$ and $A$, for generic parameters, $S$ and $B$ would be correlated (i.e. $S$ and $B$ would not be d-separated in the DAG), so that the only way to ensure $\left(S \indep B\right)$ would be by fine-tuning.  By symmetry of the independence relations under the exchange $A \leftrightarrow B$ and $S \leftrightarrow T$, one can also exclude all the possibilities wherein there is a direct causal influence from $B$ to $A$.  It follows that $A$ and $B$ must be related by a hidden common cause alone. 

We now consider the possibility of a causal influence from $A$ to $S$.  This would be an odd sort of influence, from the outcome to the setting, but we don't need to appeal to its oddness to rule it out; it can be excluded based on the principle of fine-tuning. The argument is as follows.  If such a causal influence existed, then because of the common cause acting on $A$ and $B$ (the existence of which we demonstrated above), we would find that $B$ and $S$ ought to be correlated for generic choices of the parameters (i.e. $B$ and $S$ would not be d-separated in the DAG), so that the only way to ensure $\left(S \indep B\right)$ would be fine-tuning.  By symmetry, we can also exclude the possibility of a causal influence from $B$ to $T$. 

It follows that the only causal structures that can explain the observed CI relations without fine-tuning have the following features: $A$ and $B$ have a common cause;
$S$ and $A$ are related by a direct causal influence from $S$ to $A$, or by a common cause,
 or by both mechanisms; 
 $T$ and $B$ are related by a direct causal influence from $T$ to $B$, or by a common cause,
 or by both mechanisms. 

We now demonstrate that regardless of which of the three possible causal mechanisms (depicted in Fig.~\ref{fig:XY-ambcause}) are acting between $S$ and $A$, and also regardless of which of the three are acting between $T$ and $B$, we can express the conditional probability $P(A,B|S,T)$ as follows (where $\lambda$ denotes the hidden variable which is the common cause of $A$ and $B$)
\begin{eqnarray}\label{pabst0}
P(A,B|S,T) = \sum_{\lambda}  P(A|S,\lambda) P(B|T,\lambda) P(\lambda).
\end{eqnarray}
We argue this as follows.  Begin by noting that by the definition of conditional probability, we have
\begin{eqnarray}
P(A,B,S,T) = \sum_{\lambda} P(A,S|B,T, \lambda) P(B,T|\lambda) P(\lambda).
\end{eqnarray}
However, for all nine of the causal diagrams that remain, we have $\left(AS \indep BT | \lambda \right)$, and therefore $P(A,S|B,T, \lambda)=P(A,S|\lambda)$, so that 
\begin{eqnarray}
P(A,B,S,T) = \sum_{\lambda} P(A,S| \lambda) P(B,T|\lambda) P(\lambda).
\end{eqnarray}
From this expression, we determine $P(A,B|S,T)$ using the definition of conditional probability, $P(A,B|S,T)=P(A,B,S,T)/P(S,T)$.  By assumption, $\left(S \indep T\right)$, so $P(S,T)=P(S)P(T)$ and it follows that
\begin{eqnarray}\label{pabst}
P(A,B|S,T) = \sum_{\lambda}  \frac{ P(A,S| \lambda)}{P(S)}  \frac{ P(B,T| \lambda)}{P(T)}  P(\lambda).
\end{eqnarray}
Next, we show that regardless of which of the three causal relations hold between $S$ and $A$, we have
\begin{eqnarray}\label{reducL}
\frac{ P(A,S| \lambda)}{P(S)} = P(A|S,\lambda).
\end{eqnarray}
By symmetry, this implies that regardless of which of the three causal relations hold between $T$ and $B$, we have
\begin{eqnarray}\label{reducR}
\frac{ P(B,T| \lambda)}{P(T)} = P(B|T,\lambda).
\end{eqnarray}
Eqs.~\eqref{pabst}, \eqref{reducL} and \eqref{reducR} together imply Eq.~\eqref{pabst0}.

So it remains only to prove that Eq.~\eqref{reducL} holds for each of the three possible causal relations between $S$ and $A$.  In each case, one can easily verify that $\left(S \indep \lambda\right)$.  This follows, for instance, from an application of the $d$-separation criterion (described in Appendix \ref{app:dsep}) for each of the three possible causal relations.  Recalling that this fact may be expressed as $P(S|\lambda)=P(S)$ and given that by definition $P(A|S,\lambda)= P(A,S| \lambda)/ P(S|\lambda)$, one sees that Eq.~\eqref{reducL} follows.

So we conclude that the only causal structures that can explain the observed CI relations without fine-tuning are such that $P(A,B|S,T)$ can be decomposed as in Eq.~\eqref{pabst0}.  However, it is well known that the existence of such a decomposition implies that $P(A,B|S,T)$ satisfies the Bell inequalities.  This contradicts our assumption that a Bell inequality is violated, so these causal structures are also excluded as candidate explanations of the correlations.  We have thereby exhausted the set of possible causal structures.
\end{proof}

\section{Conclusions}
\label{sec:conc}

Our two main conclusions are as follows. First, causal discovery
algorithms that appeal only to conditional independences among observed
variables cannot distinguish between Bell-inequality-violating and
Bell-inequality-satisfying correlations. Better algorithms which look to the
strength of correlations are needed to do justice to Bell's theorem.  Second, and more importantly, we have shown that any causal model which can reproduce Bell-inequality violations while respecting the observed independences---the marginal independence of the measurement settings and the no-signalling condition---will necessarily violate a principle that is at the core of all the best causal discovery algorithms, namely, that observed independences should not be explained by fine-tuning of the causal parameters in the model.  This is true in particular for all explanatory strategies that fit within the framework of directed acyclic graphs supplemented with conditional probabilities, including models that posit superluminal causes, models that exploit the superdeterminism loophole, and models that posit retrocausation.

The topic of causal discovery is still relatively young. The best algorithms
available today are not likely to be the final story.
Indeed, our analysis suggests that the tools that have been developed in
the literature on the foundations of quantum theory for assessing the possibility of \emph{local} explanations of correlations may well be important
for developing causal discovery algorithms. If one could deliver on this
promise, then it would be an interesting example of the field of quantum
foundations having applications in other fields, such as statistics and machine learning, and via these, in medicine, genetics, economics and other disciplines wherein causal discovery plays a prominent role.

Conversely, it is our view that there is a great deal more insight to be
gained about the foundations of quantum theory from the literature on causal
models and causal discovery algorithms.
We consider a few possible directions of research along these lines.

As mentioned previously,  defining causality in a manner that does not make reference to temporal ordering provides a language by which one could hope to describe a fundamental theory wherein spatio-temporal notions are emergent and notions of causal structure are primitive. In such a theory, it would not be the case that a cause was defined to be prior in time to its effects, but rather the notion of the temporal order of two events would be defined in terms of whether one event was a potential cause of the other.  Consequently, the framework for causal inference provides a natural arena in which to pursue the idea that space-time is emergent, a notion that is popular in attempts to unify general relativity with quantum theory~\cite{markopoulou2007new,dowker2005causal}.

There are a number of results in the quantum foundations literature that
have the following form: make some assumptions about the causal structure
and derive inequalities on the correlations that can be obtained from these classically.
Svetlichny's inequalities are an example of this~\cite{Svetlichny1987}, wherein one considers a
triple of measurements at space-like separation and one allows a mixture of
causal structures wherein superluminal influences can propagate between any
two of the wings of the experiment. The topic has been studied in Refs.~\cite{Mitchell2004, Acin2011,Barrett2011}.  Fritz has also recently derived inequalities on classical correlations for some causal structures that do not correspond to the standard Bell scenario~\cite{Fritz2012}.
Such results are examples of a general approach to correlations that has
been developed in the causal model literature. For instance, In Pearl's book (Sec.~8.4), inequalities on correlations are derived from assumptions about
the causal structure in a section considering noncompliance in drug trials.
Pearl points out the similarity between these ``instrumental'' inequalities and the Bell
inequalities, and adds: ``The instrumental inequality can, in a sense,
be viewed as a generalization of Bell's inequality for cases where direct
causal connection is permitted to operate between the correlated
observables, $X$ and $Y.$'' It will be interesting to see
how many results in the quantum foundations literature can be considered to
be instances of such generalized inequalities.

Finally, by exploiting a quantum analogue of conditional probability
proposed by Leifer~\cite{Leifer2006a} and developed by Leifer and Spekkens \cite{Leifer2011,Leifer2011a} and an
associated quantum analogue of conditional independence (see Leifer and
Poulin~\cite{Leifer2008}, for instance), one can hope to explore a generalization of
the notion of causal model to a \emph{quantum causal model}. A quantum
causal model is naturally defined as a quantum causal structure, which is a
directed acyclic graph wherein each node is a quantum system, and a set of
quantum causal parameters, which constitute a set of conditional quantum
states (the quantum analogue of conditional probability) for every node
given its causal parents. Insofar as one can accommodate classical
variables as special cases of quantum systems (corresponding to commuting
algebras), one can describe correlations among
settings and outcomes within quantum causal models.

Quantum causal models make similar assumptions about the possibilities
for causal structure as do classical causal models (no cycles for instance),
and they make similar assumptions about the consequences of causal structure
for statistical independences, but they replace the formalism of
classical probability theory with a noncommutative generalization thereof.
If one can make the case that the formalism of quantum causal models is not just a mathematical artifice but can be given a sensible interpretation as a form of causal explanation, then such models can provide a causal explanation of Bell-inequality violations without requiring fine-tuning.

Note, however, that if the conditional probabilities that appear in classical causal models are interpreted as degrees of belief -- and we take this to be the most sensible interpretation -- then the transition from classical causal models to quantum causal models involves not only a modification to physics, but a modification to the rules of inference. In this view, the correct theory of inference is not \emph{a priori} but empirical. Nonetheless, one cannot simply declare \emph{by fiat} that some formulation of quantum theory is a theory of inference. One must justify this claim.  At a minimum, one must determine how standard concepts in a theory of inference generalize to the quantum domain.  One could also reconsider the various proposals for axiomatic derivations of classical probability theory, for instance, that of Cox~\cite{cox1946probability} or that of de Finetti~\cite{deFinetti1937foresight}, to see whether a reasonable modification of the axioms yields a quantum theory of inference\footnote{Fuchs and Schack have also suggested that parts of quantum theory can be derived by an appeal to dutch-book coherence following de Finetti~\cite{fuchs2009quantum}.}.  Ideally, one would show that if quantum causal models imply a modification to both our physics and to our theory of inference, then these modifications are not independent.  After all, the physics determines the precise manner in which an agent can gather information about the world and in turn act upon it and so the physics should determine what is the most adaptive theory of inference for an agent. It is in this sense that the project of defining quantum causal models is not yet complete and only with such a completion in hand can one really say that a causal explanation of the Bell correlations without recourse to fine-tuning has been achieved.

\section{Acknowledgements}

RWS\ thanks Matthew Leifer for having introduced him to causal networks and for
discussions on the ideas in this article. We would also like to thank
Jonathan Barrett, Cozmin Ududec and especially Dominik Janzing for helpful discussions, and Howard Wiseman for comments on a draft of the article.  Part of
this work was completed while CJW was a student in the Perimeter Scholars
International program. \ Research at Perimeter Institute is supported in
part by the Government of Canada through NSERC and by the Province of
Ontario through MRI. 


\begin{appendix}
\section{$d$-separation}
\label{app:dsep}

\begin{figure}[h]
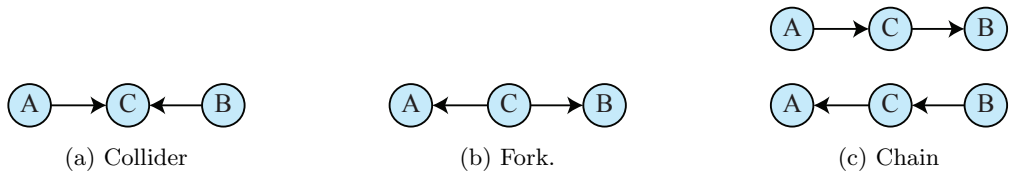

	\begin{subfigure}[b]{0.18\textwidth}
                	\centering
                \includegraphics[width=\textwidth]{a-ind-b}
             	\subcaption{Collider}
                	\label{fig:dag-collider}
        \end{subfigure}
	\hspace{5em}
        \begin{subfigure}[b]{0.18\textwidth}
                	\centering
                \includegraphics[width=\textwidth]{markovchain3}
             	\subcaption{Fork.}
                	\label{fig:dag-fork}
        \end{subfigure}
	\hspace{5em}
        \begin{subfigure}[b]{0.18\textwidth}
                	\centering
                	\includegraphics[width=\textwidth]{markovchain1}
		
		\vspace{1em}
		\includegraphics[width=\textwidth]{markovchain2}
		\subcaption{Chain}
		\label{fig:dag-chain}
	\end{subfigure}
        \caption{Basic structures found in DAGs.}
        \label{fig:dag-elts}
\end{figure}

Conditional independence relations are captured in directed acyclic graphs by the notion of distance-separation or \textbf{d-separation}. First let us introduce the basic elements of which a DAG may be composed; these are \emph{colliders}, \emph{forks}, and \emph{chains}; which for three variables $A,B,C$ are illustrated in Fig.~\ref{fig:dag-elts}.
\newline

Given a DAG $G$. A path between two vertices $X$ and $Y$ in $G$ is any set of edges and vertices which connects $X$ and $Y$, regardless of the direction of the edges. We say that a path between $X$ and $Y$ is \emph{blocked} by a set of vertices $\mathbf{Z}$ if at least one of the following conditions holds

\begin{enumerate}
\item The path contains a chain (Fig,~\ref{fig:dag-chain}), or a fork (Fig.~\ref{fig:dag-fork}) such that $C$ is in $\mathbf{Z}$.
\item The path contains a collider (Fig.~\ref{fig:dag-collider}) such that $C$ is not in $\mathbf{Z}$ and no descendant of $C$ is in $\mathbf{Z}$.
\end{enumerate}

We then have the following definition of $d$-separation:

\begin{definition}[d-separation]
Given a DAG $G$ with vertices $\textbf{V}$, two vertices $X,Y\in \mathbf{V}$ are $d$-separated by a set of vertices ${\bf Z} \subset \mathbf{V}$
  if and only if $\mathbf{Z}$ blocks all paths between $X$ and $Y$.
\end{definition}

d-separation is a relation among three sets of variables in a DAG.  If one is interpreting DAGs as causal networks (as in this article), then d-separation must represent a \emph{causal} relation among the three sets of variables.  By contrast, conditional independence represents a \emph{statistical} relation among them.  One might say that $\mathbf{X}$ is \emph{causally screened off} from $\mathbf{Y}$ given $\mathbf{Z}$ whenever $\mathbf{X}$ is d-separated from $\mathbf{Y}$ given $\mathbf{Z}$.  Of course, the significance of this causal relation is found in the statistical distributions that can be supported by the causal structure.  A set of variables $\mathbf{X}$ is d-separated from the set $\mathbf{Y}$ given the set $\mathbf{Z}$ in a causal structure if and only if for all probability distributions over the causal structure, $\mathbf{X}$ is conditionally independent of $\mathbf{Y}$ given $\mathbf{Z}$.

\end{appendix}

\end{document}